\documentclass[aps, pra,twocolumn,english,reprint,longbibliography,superscriptaddress,breaklinks=true,showkeys,showpacs=false,nofootinbib]{revtex4-2}

\usepackage[T1]{fontenc}
\usepackage[utf8]{inputenc}
\usepackage{color}
\usepackage{tikz}

\usepackage{amsmath}
\usepackage{amssymb}
\usepackage{graphicx}
\usepackage{physics} 
\usepackage{dsfont}
\usepackage{hyperref}
\usepackage{amssymb}
\usepackage{svg}
\usepackage{bbold} 
\usepackage{amsthm}
\newtheorem{definition}{Definition}
\newtheorem{proposition}{Proposition}
\newtheorem{theorem}{Theorem}
\usepackage{subfig}
\usepackage{ragged2e}

\begin{document}

\title{Quantum thermodynamics as a gauge theory}

\author{Gabriel Fernandez Ferrari}
\email{gabrielferrari@discente.ufg.br}
\affiliation{QPequi Group, Institute of Physics, Federal University of Goi\'as, Goi\^ania, Goi\'as, 74.690-900, Brazil}

\author{\L ukasz Rudnicki}
\email{lukasz.rudnicki@ug.edu.pl}
\affiliation{International Centre for Theory of Quantum Technologies (ICTQT), University of Gda\'{n}sk, 80-308 Gda\'{n}sk, Poland}

\author{Lucas C. C\'eleri}
\email{lucas@qpequi.com}
\affiliation{QPequi Group, Institute of Physics, Federal University of Goi\'as, Goi\^ania, Goi\'as, 74.690-900, Brazil}

\begin{abstract}
Thermodynamics is based on a coarse-grained approach, from which its fundamental variables emerge, effectively erasing the complicated details of the microscopic dynamics within a macroscopic system. The strength of Thermodynamics lies in the universality provided by this paradigm. In contrast, quantum mechanics focuses on describing the dynamics of microscopic systems, aiming to make predictions about experiments we perform, a goal shared by all fundamental physical theories, which are often framed as gauge theories in modern physics. Recently, a gauge theory for quantum thermodynamics was introduced, defining gauge invariant work and heat, and exploring their connections to quantum phenomena. In this work, we extend that theory in two significant ways. First, we incorporate energy spectrum degeneracies, which were previously overlooked. Additionally, we define gauge-invariant entropy, exploring its properties and connections to other physical and informational quantities. This results in a complete framework for quantum thermodynamics grounded in the principle of gauge invariance. To demonstrate some implications of this theory, we apply it to well-known critical systems.
\end{abstract}

\maketitle

\section{Introduction}
    Thermodynamics has been referred to as the witch of all physical theories~\cite{Goold2016} due to its distinct nature among physical models. While fundamental theories like General Relativity and the Standard Model aim to describe how processes unfold, thermodynamics focuses on which processes are possible and which are not allowed to occur in nature\footnote{Reference.~\cite{Goold2016} starts with the following sentence: \emph{If physical theories were people, thermodynamics would be the village witch. Over the course of three centuries, she smiled quietly as other theories rose and withered, surviving major revolutions in physics, like the advent of general relativity and quantum mechanics. The other theories find her somewhat odd, somehow different in nature from the rest, yet everyone comes to her for advice, and no-one dares to contradict her.}}. Classically, thermodynamics consists of three laws. The zeroth law defines the equilibrium state, the first law states the conservation of energy, while the second one gives us the direction in which physical processes can occur, which is the direction pointing to entropy growth. Quantum mechanical considerations lead us to the fourth law, which states the unattainability of absolute zero temperature~\cite{Callen1985}.

The universality of Thermodynamics lies in the nature of the questions it addresses. In the thermodynamic limit, our measurements of space and time intervals capture only the averages of microscopic dynamics, as our instruments cannot see the tiny atomic scales. This coarse-grained approach gives rise to the thermodynamic variables, leading to a strong theory that limits all physical processes, regardless of the laws that apply to the microscopic degrees of freedom of a macroscopic system. Such a feature turns possible the application of this theory to several fields, ranging from engineering~\cite{Bella2021} to biology~\cite{Mizraji2024} and neuroscience~\cite{Deli2021}, passing through complex systems~\cite{Gross2001}, computation~\cite{Bennett1982} and black holes~\cite{Wald2001}, just to mention a few.

However, there is a special feature of equilibrium Thermodynamics. In the limit where it was developed, the thermodynamic limit, fluctuations vanish. When we consider small systems, or quantum systems, the role of fluctuations must be taken into account. Also, when the system is taken out of equilibrium, the usual formulation of Thermodynamics does not apply. In these cases, we must employ stochastic thermodynamics~\cite{Strasberg2022}, which is a well-developed subject, specially in classical physics. There are also other approaches to face this problem, with information theory~\cite{Goold2016} and axiomatic formulations~\cite{Lieb1999,Giles2016} standing among the most known ones.

Although significant progress has been made since Alicki's seminal work~\cite{Alicki1979}, the theory of Quantum Thermodynamics~\cite{Alicki2018,Binder2018} is still not fully developed. Many open questions remain, particularly regarding the fundamental definitions of key thermodynamic quantities: work, heat, and the thermodynamic entropy. 

In this context, work is typically defined as the energy change in a system due to an external, time-dependent process that changes the Hamiltonian of the system. One common approach is the two-point measurement scheme, where work is defined as the difference in energy between two projective measurements of the Hamiltonian at the beginning and end of the process~\cite{Kurchan2000,Talkner2007}. This definition captures the stochastic nature of work in quantum systems due to the unavoidable quantum fluctuations. Another perspective, based on quantum trajectories and the Stochastic Thermodynamics framework, defines work through the averaged outcomes of individual quantum trajectories~\cite{Horowitz2012}. Additionally, there are definitions involving quantum coherence, which highlight the role of off-diagonal elements in the density matrix and their contribution to the work distribution in coherent quantum processes~\cite{Lostaglio2015}. See Ref.~\cite{Talkner2016} for a broad perspective on quantum work.

The definition of thermodynamic entropy faces similar problems to be extended to the quantum realm. The most commonly used definition is the von Neumann entropy~\cite{Neumann1927}. However, being a measure of the information content of a quantum state, it is fundamentally distinct from the thermodynamic entropy. Another important concept is the diagonal entropy, which considers only the diagonal elements of the density matrix in the energy eigenbasis of the system thus reflecting classical probabilistic distributions~\cite{Polkovnikov2011}. Additionally, the concept of relative entropy, which quantifies the distinguishability between two quantum states, is important for understanding irreversibility and the arrow of time in quantum processes~\cite{Kawai2007,Sagawa2013,Vedral2002}. Despite all the controversies, these definitions collectively help to extend thermodynamic principles to quantum systems, allowing for a rigorous analysis of entropy production, coherence, and the fundamental limits of quantum thermodynamic processes~\cite{Landi2021}.

Taking a completely distinct path, fundamental physical theories aim to answer questions about scattering amplitudes and the trajectories of planets and galaxies. These models seek a comprehensive description of physical systems, including the dynamics of the microscopic degrees of freedom. In their modern formulation, these theories are expressed as gauge theories, which are a framework providing a description of the fundamental interactions through symmetry. At their core, gauge theories postulate that the laws of physics (and all physical quantities) should be invariant under certain local transformations, known as gauge transformations. The most prominent example is the Standard Model of particle physics, which describes electromagnetic, weak, and strong interactions using gauge theories based on the special unitary group~\cite{Weinberg1995}. Additionally, gauge theories have profound implications in General Relativity, where the gravitational interaction is described by the gauge theory of the Poincaré group, reflecting the invariance under local Lorentz transformations and translations~\cite{Rovelli2004}. The description of the laws of physics in terms of gauge theories not only provides a consistent theoretical framework but also drives the search for a deeper understanding of fundamental interactions and the development of new physics.

Recently, Quantum Thermodynamics was formulated as a gauge theory of the Thermodynamic Gauge Group~\cite{gauge-lucas}. This approach is based on the fundamentally distinct paradigms of classical Thermodynamics and Quantum Mechanics. In classical Thermodynamics, all quantities are derived from a coarse-graining procedure, whereas in Quantum Mechanics, we assume control over all microscopic degrees of freedom, knowing the quantum state of the system, which contains too much information from a thermodynamic perspective.

The theory proposed in Ref.~\cite{gauge-lucas} defines a symmetry group called the thermodynamic gauge group, or thermodynamic group, which implements a form of coarse-graining in the information space. Just as the microscopic dynamics of a classical system can be considered redundant for Thermodynamics, part of the information in the quantum state is redundant for Quantum Thermodynamics. The thermodynamic group was introduced exactly to eliminate this redundancy. We therefore have described Quantum Thermodynamics as a gauge theory of the thermodynamic group. By applying the gauge invariance principle, the quantum thermodynamic quantities uniquely emerge from this formalism. In Ref.~\cite{gauge-lucas} the gauge-invariant notions of work and heat, as well as their connections to quantum features like quantum coherences was extensively discussed.

The main results of the present work can be summarised as follows. First, we extend the theory presented in Ref.~\cite{gauge-lucas} by explicitly incorporating systems with degenerate energy eigenvalues, which provides new insights into the definitions of work and heat in the quantum domain. Secondly, we introduce the concept of gauge-invariant entropy, which accounts for all quantum features, including degeneracies and quantum coherences. Our findings suggest that the gauge theory of Thermodynamics may offer a unified description of quantum and classical thermodynamics. More importantly, this is a robust and consistent framework that is grounded in a fundamental physical principle: gauge invariance. 

There is an important distinction between the thermodynamic group and the gauge groups associated with other physical theories. Fundamental physical theories are described by fields, with the associated potentials being the gauge fields. While fields are the crucial physical quantities, the potentials lack physical relevance due to the gauge freedom in their definition. The role of the gauge invariance principle is precisely to eliminate this redundant information, transitioning from potentials to physical quantities such as scattering amplitudes, establishing a fundamental symmetry of the system. In our context, the density matrices, the carriers of information, plays the role of the potentials. Consequently, the information within the system's state is redundant only from a thermodynamic perspective, but remains significant in Quantum Information or Quantum Mechanics. Thus, the thermodynamic group is not a fundamental symmetry of nature but reflects our inability to fully control the system. For instance, in experiments with many energy levels, quantum tomography is impractical, and we typically have access only to energy measurements~\cite{Araujo2018}. Therefore, the state information about the basis is irrelevant for computing thermodynamic quantities. The thermodynamic group was proposed to consistently eliminate such information. Therefore, the role of he thermodynamic group is to take us from the space of states to the thermodynamic variables by removing redundant information, implementing a coarse-graining akin to classical thermodynamics.

This remaining of the paper is organised as follows. In Sec.~\ref{sec:gauge} we present the formalism of gauge theory of Quantum Thermodynamics. Beyond reviewing the theory proposed in Ref.~\cite{gauge-lucas}, we generalise it to the case where the Hamiltonian spectrum is degenerate and its degeneracies change over time. Next, in Sec.~\ref{sec:laws} we discuss how the general expressions for heat and work fits the first law of quantum thermodynamics. Moreover, this section also provides the definition of the gauge invariant entropy and a discussion regarding the second law. A comparison between the invariant thermodynamic quantities with the most known proposals in literature is also presented. We then apply our formalism to quantum critical systems in Sec.~\ref{sec:transitions}. Finally, a deeper physical analysis of our results is presented in Sec.~\ref{sec:discussions}.

\section{Gauge theory of Quantum Thermodynamics}
\label{sec:gauge}

\indent We begin by defining the thermodynamic gauge group, initially introduced in Ref.~\cite{gauge-lucas}. We advance the theory by first extending it to systems with degenerate energy eigenvalues. Following this, we provide the general definition of physical quantities that satisfy the gauge invariance postulate, including the invariant entropy. This is the gauge theory of the thermodynamic group. With these fundamental quantities in hand, we then discuss both the first and second laws of Quantum Thermodynamics as they emerge from the proposed framework.

\subsection{Thermodynamic group}

\indent Let us consider a quantum system associated with the Hilbert space $\mathcal{H}^{d}$, whose dimension is denoted as
$\dim\{\mathcal{H}^{d}\}=d$. To construct the thermodynamic group, we assume the following: (i) $\mathcal{H}^{d}$ is finite-dimensional; (ii) the thermodynamic quantities are functionals of the density matrix; (iii) the gauge transformations are unitary.

The first assumption is not overly restrictive since $d$ can be as large as we please. It is important to remember that most experiments involve a limited number of particles, typically under well-defined boundary conditions. Therefore, we usually deal with systems that effectively have a countable number of energy eigenstates. Even for a continuous degree of freedom, detectors have finite resolution, which introduces coarse-graining into the system, thus allowing for a description in terms of finite, countable energy states. The second assumption asserts that every physical quantity must be a functional of the density operator, which contains the information about the system. While it can also be a function of other operators, such as the Hamiltonian, it will always be a function of the density matrix. The third assumption is based on the fact that every continuous symmetry group, which is the case here, admits a unitary representation. We will revisit this point at the end of the paper. 

These assumptions alone are not sufficient to define the thermodynamic group. Considering all possible unitary transformations would render the theory trivial and not useful~\cite{gauge-lucas}. Therefore, we must impose a physical restriction on this set. Since we are discussing Thermodynamics and symmetry transformations, it is natural to require that the action of the gauge group should not change the mean energy of the system
\begin{equation}
    U[\rho] = \Tr\left\{\rho H_t\right\}, \label{energy-mean}
\end{equation}
with $H_t$ and $\rho$ being the (possibly time-dependent) system Hamiltonian and density matrix, respectively. From this, we define the following principle
\begin{definition}
\label{Emergent thermodynamic gauge}
    (Gauge invariance of energy~\cite{gauge-lucas}) An unitary transformations $V_t:\mathcal{H}^d\to \mathcal{H}^d$ are an admissible thermodynamic gauge transformation if it preserves the mean energy
    \begin{equation}
    U[V_t \rho V_t^{\dagger}] = U[\rho] \label{Emg}
   \end{equation}
    for all density matrices $\rho$.
\end{definition}
\indent Definition~\ref{Emergent thermodynamic gauge} establishes the invariance principle under the thermodynamic group and defines the time-dependent gauge transformation $V_t$. It is important to note that this transformation is not an actual process acting on the system but rather a symmetry transformation with no physical consequences. This is the essence of a gauge symmetry.

Using this definition, we can explicitly construct the thermodynamic group~\cite{gauge-lucas}. First, note that energy invariance under $V_t$ implies that $[V_t,H_t] = [V^{\dagger}_t,H_t] = 0$, since this must hold for any density operator. Therefore, the gauge transformations $V_t$ are unitary channels that commute with the Hamiltonian. It is important to emphasise that $V_t$ is not equivalent to the thermal operations defined in the context of the resource theory of thermodynamics~\cite{Brandao2013,Chitambar2019}. Instead, the thermodynamic gauge forms a very particular subclass of these transformations, which are not associated with any physical interpretation. For more details, see Ref.~\cite{gauge-lucas}.

The construction of the symmetry group proceeds as follows. Since the Hamiltonian is hermitian, it can be written as
\begin{equation}
        H_t = u_t \left(\bigoplus_{k=1}^{p} \lambda_{t}^{k} \mathbb{1}_{n_{t}^{k}}\right)u_t^{\dagger} = u_t h_t u_t^{\dagger}, \label{H_t decomposition}
\end{equation}
where $p$ is the number of distinct energy eigenvalues $\lambda^{k}_t$, while $n_t^k$ stands for the multiplicity of each $\lambda^{k}_t$, at each instant of time $t$. That is, $n_t^k$ is the instantaneous degree of degeneracy of the $k$-th energy eigenvalue. We must have $\sum_{k=1}^{p}n_t^k = d$. $\mathbb{1}_{a}$ is the identity matrix with dimension $a$. Note that $n_{t}^{k}$ is allowed to change in time. This will have very important physical consequences.

Given the Hamiltonian decomposition in Eq.~\eqref{H_t decomposition}, along with the condition $[H_t,V_t]=0$, all $V_t$ must be of the form
   \begin{equation}
        V_t = u_t \mathcal{V}_{t} u_t^{\dagger}, \hspace{0.3cm} \mbox{with} \hspace{0.3cm}\mathcal{V}_{t} = \bigoplus_{k=1}^{p} v^k_t  ,\label{unitary-gauge-transformation}
  \end{equation}
where $v^k_t \in \mathcal{U}(n_t^k)$, the set of unitary matrices of dimension $n_t^k$. From the perspective of group representation theory, $\mathcal{V}_{t}$ is decomposed in unitary irreducible representations $v_t^k$ of each $\mathcal{U}(n_t^k)$. Importantly, this is associated to the decomposition of the Hilbert space $\mathcal{H}^d = \bigoplus_{k=1}^{p}\mathcal{H}_k$ where each $\mathcal{H}_k$ is the subspace spanned by the eigenvectors associated to the $k$-th eigenvalue.

Furthermore, Eq.~\eqref{unitary-gauge-transformation} provides another characterisation to the thermodynamic group. In fact, $[V_t, H_t]~=~0~\iff~[\mathcal{V}_{t},h_t]~=~0$ and we can describe the gauge transformation using both $V_t$ or $\mathcal{V}_{t}$. Therefore, regarding Definition~\ref{Emergent thermodynamic gauge}, $\mathcal{V}_{t}$ preserves the mean energy with respect to diagonal Hamiltonian $h_t$.

Now, from Eq.~\eqref{unitary-gauge-transformation} we can formally define
\begin{definition}
\label{instantaneos emergent thermodynamic gauge group}
    (Thermodynamic group) Let $H_t$ be a time dependent Hamiltonian operator defined on a $d$-dimensional Hilbert space $\mathcal{H}^d$, and $u_t$ be defined as in Eq.~\eqref{H_t decomposition}. Then, the thermodynamic group is defined to be the following set of transformations
        \begin{align}
        \mathcal{T}_{H_t}=\Big\{ V_t \in \mathcal{U}(d) |& \left[V_t,H_t\right] = 0, V_t= \left.u_t\left(\bigoplus_{k=1}^{p} v^k_t\right) u_t^{\dagger} \right\} \label{Ins-emergent-gauge-group1}
    \end{align}
    where  $v^k_t \in \mathcal{U}(n_t^k) \subset \mathcal{U}(d)$ and  $\Gamma  = \{n_t^k\}_{k=1,...,p}$ is the set of labels for the degeneracies of the eigenvalues of $H_t$ at time $t$.
\end{definition}

The topological aspects of the thermodynamic group can be analysed by introducing the isomorphic group $\mathcal{G}_{\operatorname{T}}$, with is associated to each matrix $\mathcal{V}_{t}$ in Eq.~\eqref{unitary-gauge-transformation}. 
\begin{proposition}
\label{isomorphism-props}
($\mathcal{G}_{\operatorname{T}}$-group) The thermodynamic group is isomorphic to
    \begin{equation}
    \mathcal{G}_{\operatorname{T}}  = \mathcal{U}(n^1_t) \times \mathcal{U}(n_t^2) \times ... \times \mathcal{U}(n_t^k), \label{isomorphism}
    \end{equation}
    at each instant of time $t$. The symbol $\times$ denotes the Cartesian product.
\end{proposition}
Given such an isomorphism, from now on we refer to $\mathcal{G}_{\operatorname{T}}$ as the thermodynamic group.

\begin{figure}[ht!]
    \centering
    \includegraphics[scale=0.5]{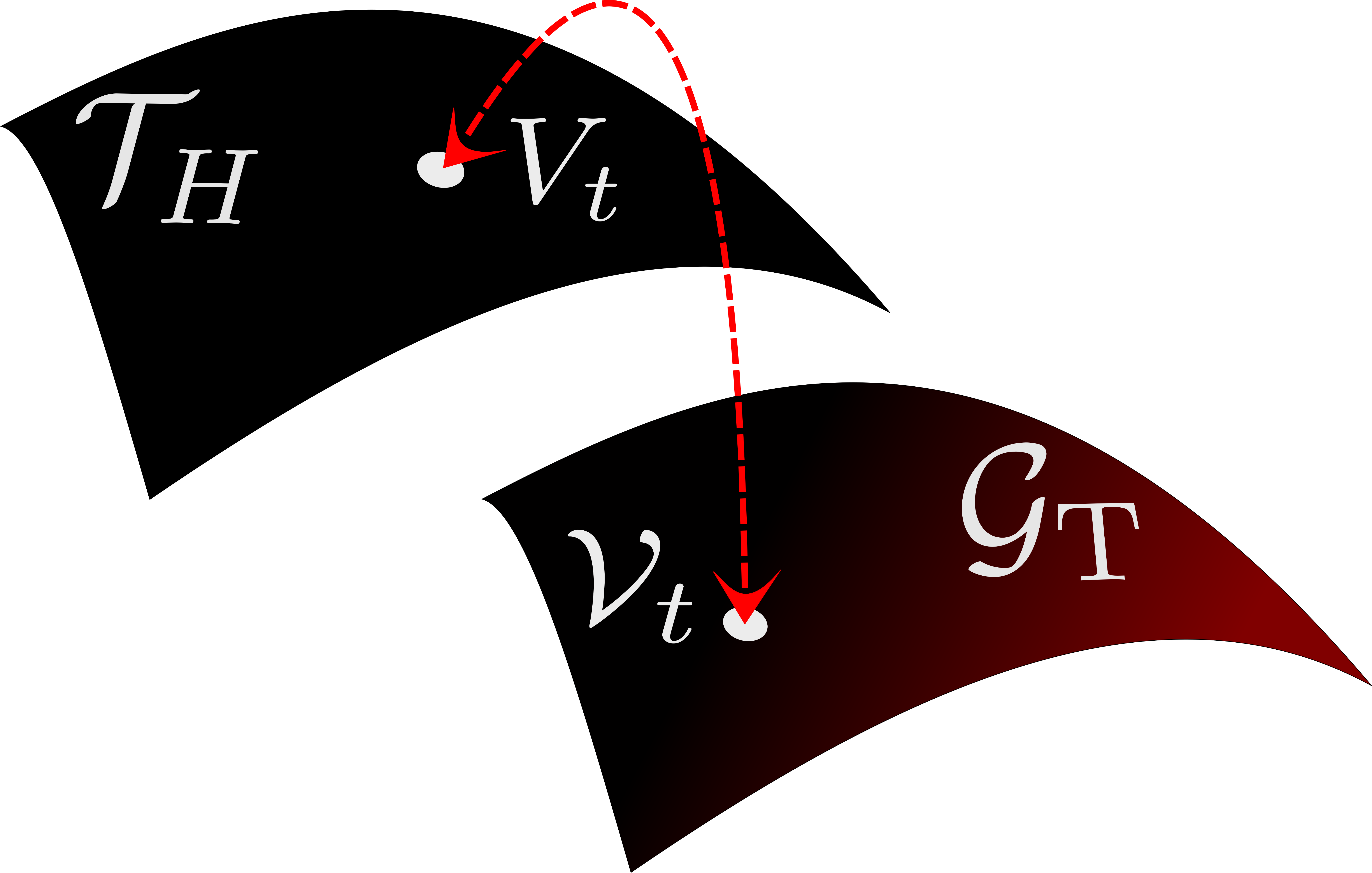}
    \caption{ \justifying Representation of the isomorphism between the thermodynamic group $\mathcal{T}_{H_t}$ and the group $\mathcal{G}_{\operatorname{T}}$.}
    \label{gauge-manifold}
\end{figure}

The group $\mathcal{G}_{\operatorname{T}}$ is directly associated with each matrix $v^k_t$ in Eq.~\eqref{unitary-gauge-transformation} and the subgroups $\mathcal{U}(n_t^k)$ of $\mathcal{U}(d)$, this is a schematic representation in Fig.~\ref{gauge-manifold}. From the topological point of view, the thermodynamic group is constructed from a product of finite compact Lie subgroups. $\mathcal{G}_{\operatorname{T}}$ is therefore, also a compact Lie group, whose Lie algebra $\mathfrak{g}_{\operatorname{T}}$ fulfills the relation $\dim\{\mathfrak{g}_{\operatorname{T}}\} \leq \dim\{\mathfrak{u}(d)\}$, which follows from Eq.~\eqref{isomorphism} as a consequence of degeneracy.

In this context, consider that if a given eigenvalue $\lambda^{m}_{t}$ is non-degenerate, the corresponding matrix $v^{m}_t$ satisfies $v^{m}_t \in \mathcal{U}(1)$. Consequently, $v^{m}_t$ contributes to the matrix $\mathcal{V}_{t}$ with only one arbitrary element, namely $(\mathcal{V}_{t})_{mm}$, while all other elements associated with this block vanish, i.e., $(\mathcal{V}_{t})_{mj}=(\mathcal{V}_{t})_{jm} = 0$ for $j\neq m$. In contrast, if any energy eigenvalue is degenerate, the associated matrix $v^k_t$ is not necessarily diagonal. In fact, the most general form of $v^{k}_t$ is a unitary matrix of order $n_t^k\times n_t^k$, i.e. $v^{k}_{t}\in \mathcal{U}(n_t^k)$, with all elements non-zero.

Therefore, each non-degenerate eigenvalue contributes to the dimension of the Lie algebra of the thermodynamic group with a basis associated with the Lie algebra of the group $\mathcal{U}(n_{t}^{k}=1)$, whose dimension is $1$. On the other hand, degenerate eigenvalues contribute to the dimension of the Lie algebra with $\mathcal{U}(n_t^k>1)$, whose dimension is $(n_t^k)^2$.

It is interesting to consider two illustrative situations. The first case, explored in Ref.~\cite{gauge-lucas}, corresponds to a fully non-degenerate energy spectrum, i.e. $n_t^k=1$ for all $k$. In this case, $\mathcal{V}_{t}$ is diagonal, since $\left[\mathcal{V}_{t},h_t\right] = 0$ and $h_t$ are diagonal with distinct elements. Therefore, the dimension of the Lie algebra associated with $\mathcal{G}_{\operatorname{T}}$ is $\dim\{\mathfrak{g}_{\operatorname{T}}\} = d$, and the group is a $d$-fold tensor product of trivial phases which are elements of $\mathcal{U}(1)$. The second interesting case involves a Hamiltonian with a fully degenerate spectrum. In this case, the gauge group is $\mathcal{G}_{\operatorname{T}} = \mathcal{U}(d)$, and the dimension of the associated Lie algebra is $d^2$. The group is a $d^2$-fold tensor product of elements belonging to $\mathcal{U}(d)$.

Now that we have the structure of the thermodynamic group, we can use the gauge invariance principle to construct all the quantities physically relevant in our theory. Such quantities are those that remain invariant under the action of $\mathcal{G}_{\operatorname{T}}$.


\section{\texorpdfstring{Gauge theory of the $\mathcal{G}_{\operatorname{T}}$-group}{gt-entropy and diagonal entropy}}

\indent\indent  The preceding analysis provides insights into the behaviour associated with the gauge transformations. Notably, these findings will be relevant in the subsequent discussion when we introduce the concept of gauge-invariant quantities with respect to the thermodynamic group. Before proceeding, it is important to note that since Eq.~\eqref{isomorphism} represents a compact Lie group~\cite{representation}, there exists a unique, normalised left and right invariant measure~\cite{representation,haar-information} associated with the group $\mathcal{G}_{\operatorname{T}}$, which we denote as
\begin{equation}
    \dd\mathcal{G}_{\operatorname{T}} =  \dd\mu\left[\mathcal{U}(n^1_t)\right] \cdot \dd\mu\left[\mathcal{U}(n^2_t)\right]  \cdot ... \cdot \dd\mu\left[\mathcal{U}(n^k_t) \right]\label{haar-measure}
\end{equation}
where $\dd\mu\left[\mathcal{U}(n^k_t)\right]$ are the Haar measures associated with the unitary group $\mathcal{U}(n_t^k)$ for  all values of $k$. $\dd \mathcal{G}_{\operatorname{T}}$ is therefore a multidimensional Haar measure. Note that the induced measure on $\mathcal{G}_{\operatorname{T}}$ in Eq.~\eqref{haar-measure} is not static. This structure evolves in time alongside the changes in the degeneracies of the Hamiltonian $H_t$. 

From Eq.~\eqref{haar-measure}, we can define the notion of gauge-invariant quantities with respect to $\mathcal{G}_{\operatorname{T}}$ using group averaging techniques.
\begin{definition}
\label{quantities}
    (Physical quantities). Given a physical quantity represented by a functional $F[\rho]$, its counterpart which is invariant with respect to the thermodynamic group $\mathcal{G}_{\operatorname{T}}$ is given by
    \begin{equation}
        F_{inv}[\rho] = \int \dd\mathcal{G}_{\operatorname{T}} F\left[V_t \rho V_t^{\dagger} \right].
        \label{invariant-1}
    \end{equation}
    If $F[\rho]$ is invariant under unitary transformations, then its gauge-invariant version is given by \begin{equation}
        F_{inv}[\rho] = F\left[\operatorname{D}(\rho)\right], \label{invariant-2}
   \end{equation}
    where $\operatorname{D}(\cdot) \equiv \int \dd\mathcal{G}_{\operatorname{T}} V_t (\cdot) V_t^{\dagger}$. $V_t$ and $\dd\mathcal{G}_{\operatorname{T}}$ are given in Eqs.~\eqref{unitary-gauge-transformation} and~\eqref{haar-measure}, respectively.  
\end{definition}

\begin{figure}[ht!]
    \centering
    \includegraphics[scale=0.71]{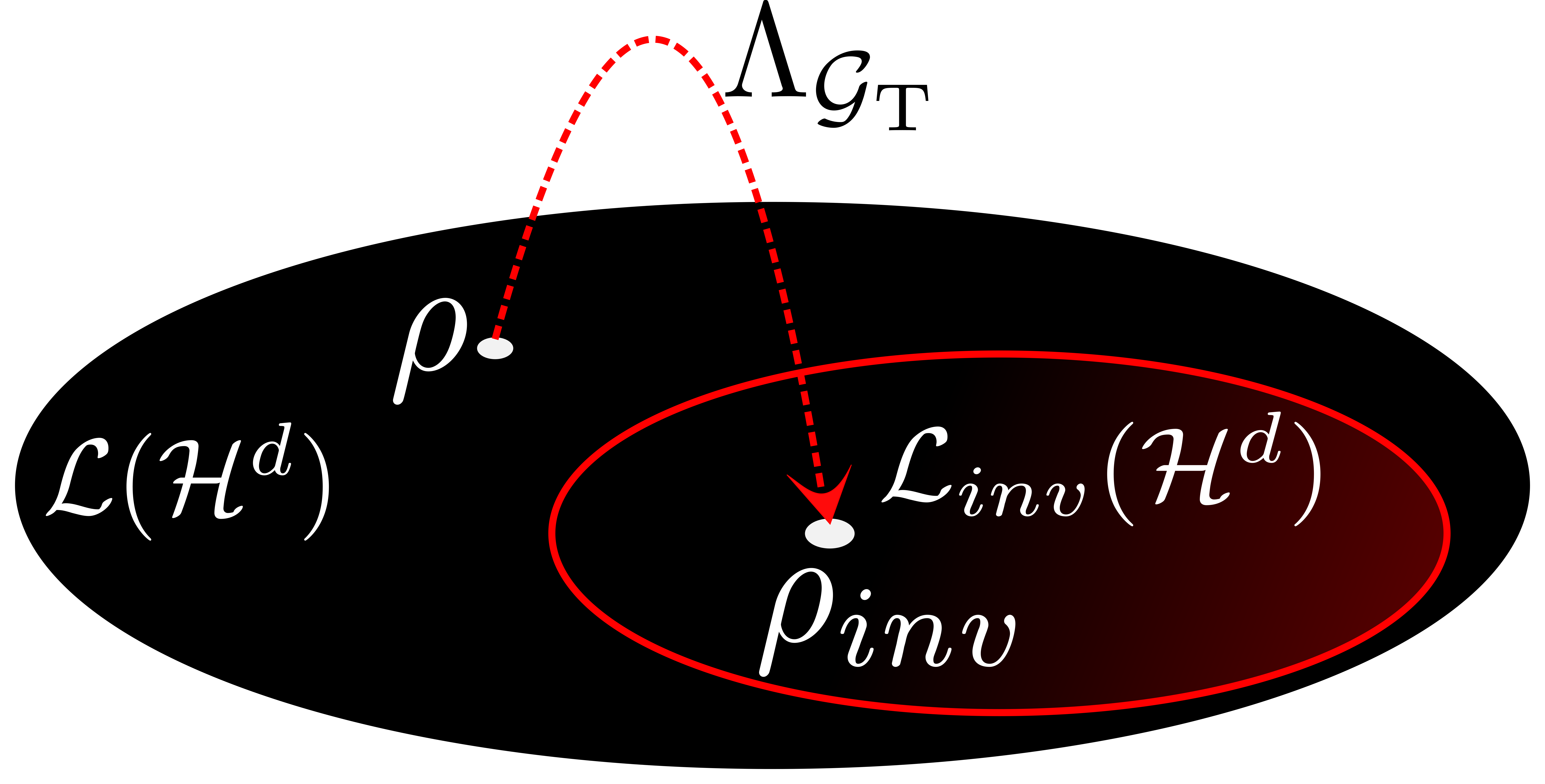}
    \caption{\justifying Schematic representation of the action of the map $\Lambda_{\mathcal{G}{\operatorname{T}}}$ on density operators in the space $\mathcal{L}(\mathcal{H}^d)$, focusing on operators that are invariant under thermodynamic group transformations. These operators are defined within the subspace of invariant operators, $\mathcal{L}_{inv}(\mathcal{H}^d) \subset \mathcal{L}(\mathcal{H}^d)$. Note that the map $\Lambda_{\mathcal{G}_{\operatorname{T}}}$ always maps the states $\rho$ to a subspace of lower dimension. Physically, this is associated with the Coarse-Graining process induced by the thermodynamic gauge group $\mathcal{G}_{\operatorname{T}}$, which eliminates information that is redundant from a thermodynamic perspective.  }
    \label{inv-space}
\end{figure}

Some words on the two quantities appearing in Definition~\ref{quantities} are in order here. As a consequence of the mathematical definition of gauge transformation assumed in this work, they are necessary. The aim of a gauge transformation is to wash away redundant information contained in quantum states. However, given the structure of $\mathcal{G}_{\operatorname{T}}$, any physical quantity that is invariant under unitary transformations is immediately invariant under the Haar integral given in Eq.~\ref{invariant-1}. Therefore, if we take only Eq.~\eqref{invariant-1}, we obtain $F_{inv}[\rho] = F[\rho]$ and the redundancy is not removed. Then, we introduce Eq.~\eqref{invariant-2} as physical quantities for the special case in which the functional $F[\rho]$ is invariant under unitary transformations, so we can indeed eliminate  redundant information. Moreover, in this case we can express the linear operator $\operatorname{D}$ using  Eq.~\eqref{unitary-gauge-transformation} as
\begin{equation}
    \operatorname{D}(\rho) =  u_t \left( \int \dd\mathcal{G}_{\operatorname{T}} \mathcal{V}_{t} \rho^E \mathcal{V}_{t}^{\dagger}\right)u_t^{\dagger}, \label{linear-D}
\end{equation}
where $\rho^{E} = u_t^{\dagger}\rho u_t$ corresponds the density matrix in the energy eigenbasis (denoted by superscript $E$). The term appearing within parentheses can be identified as a superoperator commonly referred to as a quantum twirling operator, $\Lambda_{\mathcal{G}_{\operatorname{T}}}(\cdot) \equiv \int d \mathcal{G}_{\operatorname{T}} \mathcal{V}_{t}\cdot\mathcal{V}_{t}^{\dagger}$ \cite{quantumt-twirling1}, which is a unital quantum channel~\cite{quantum-channel,nielsen}.he map $\Lambda_{\mathcal{G}{\operatorname{T}}}$, whose action is schematized in Fig.~\ref{inv-space}, acts on the space of linear operators $\mathcal{L}(\mathcal{H}^d)$, transforming the states $\rho$ into invariant operators associated with what we call the invariant subspace $\mathcal{L}_{inv}(\mathcal{H}^d)$. This action corresponds to the Haar average of the density matrices, in the energy basis, over the thermodynamic gauge group.

This average can be evaluated using group average techniques, yielding the following expression
\begin{equation}
    \rho_{dd}^E(t) \equiv \Lambda_{\mathcal{G}_{\operatorname{T}}}[\rho^E(t)]  =  \bigoplus_{k=1}^{ p\leq d} \dfrac{\operatorname{Tr} \left\{\rho^E_{n_t^k}(t)\right \}}{n_t^k} \mathbb{1}_{n_t^k}, \label{rhodd}
\end{equation}
where $\rho^E_{n_t^k}\equiv\Pi_{n_t^k}\rho(t)\Pi_{n_t^k}$, with $\Pi_{n_t^k}$, is the projector associated to each subspace $\mathcal{H}_k$ and the subscript $dd$ in $\rho_{dd}^E$ labels the diagonal and degenerate density matrix. We present some details of the Haar averaging in Appendix~\ref{Appendix-A}.

In this sense, this result implies that $F_{inv}[\rho(t)] = F\left[\Lambda_{\mathcal{G}_{\operatorname{T}}}(\rho^E(t))\right]=F\left[\rho_{dd}^E(t)\right]$, since $F$ is invariant under unitary transformations. A special point in this context is that, since the superoperator $\operatorname{D}$, or equivalently quantum twirling $\Lambda_{\mathcal{G}_{\operatorname{T}}}$, transforms the density matrix to the energy eigenbasis, all redundant information has been eliminated. 

Lastly, it is necessary to comment on two distinct points regarding this framework. First, due to the constraint established by Definition~\ref{Emergent thermodynamic gauge}, there is a natural class of quantities that are gauge-invariant by construction. This includes quantities solely defined on the equilibrium manifold or possibly on a set of instantaneous equilibrium states. More generally, all quantities depend only on the Hamiltonian. For example, the microcanonical state is preserved~\cite{gauge-lucas}.

The second point pertains to the existence of two types of transformations in this framework. The first set comprises the unitary transformations that are elements of $\mathcal{G}_{\operatorname{T}}$, denoted as $V_{t}$. These transformations describe symmetry operations on the system rather than dynamical processes. Their role is to identify the set of states, at each instant of time, that cannot be distinguished by measuring thermodynamic variables. The second set of transformations is associated with physical processes and can be any completely positive and trace-preserving map. These transformations govern the time evolution of the physical system. Therefore, the theory applies to any allowed quantum dynamics~\cite{gauge-lucas}.


\section{The laws of quantum thermodynamics}
\label{sec:laws}

Now that we have defined physical quantities, we can discuss the first and second laws of Quantum Thermodynamics. First, we extend the definitions of work and heat introduced in Ref.~\cite{gauge-lucas} to the case where the energy spectrum is degenerate. This extension leads us to the generalized first law. Next, we define the thermodynamic entropy that is gauge invariant, which enables us to state the second law of Quantum Thermodynamics.

\subsection{Work, heat and the first law}

\indent As previously noted, all physical quantities relevant to thermodynamics are represented as functionals of the system's density matrix, $\rho_t$. Specifically, we often focus on a particular set of physical quantities, such as energy, heat, and work.

Usually, work is defined as
\begin{equation}
    W_{u}[\rho] =  \int_{0}^{\tau}\dd t \Tr\left\{\rho \dv{H_t}{t} \right\} \label{usual-work}
\end{equation}
while heat takes the form
\begin{equation}
    Q_{u}[\rho] =  \int_{0}^{\tau}\dd t \Tr\left\{\dv{\rho}{t} H_t\right\}. \label{usual-heat}
\end{equation}
\indent \indent The subscript $u$ denotes the conventional definitions of thermodynamic quantities, as introduced in Ref.~\cite{Alicki1979}. This distinction is necessary to differentiate them from the quantities arising from the gauge theory we present here. These expressions form the fundamental quantities in the context of the first law of thermodynamics, which states the conservation of energy: $Q_u = \Delta U - W_u$, where $\Delta U$ represents the change in the internal energy of the system.

The application of Definition~\ref{quantities} to the usual notions of work [Eq.~\eqref{usual-work}] and heat [Eq.~\eqref{usual-heat}] provided the notions of work and heat which are invariant with respect to the thermodynamic group~\cite{gauge-lucas}. These are the physical work and heat we refer in the present framework.
\begin{theorem}
\label{theo-lucas}
    (Gauge-Invariant work and heat~\cite{gauge-lucas}). Let $H_t$ be a possibly time-dependent hamiltonian, whose decomposition is given in Eq.~\eqref{H_t decomposition}, and $u_t$ the unitary transformation that diagonalizes $H_t$, i.e $H_t = u_t h_t u_t^{\dagger}$, which is differentiable for all $t \in \mathbb{R}$. Then, the notion of work and heat which is invariant with respect to the thermodynamic group $\mathcal{G}_{\operatorname{T}}$ are given by
    \begin{equation}
W_{inv}\left[\rho\right]=\int_{0}^{\tau} \dd t \operatorname{Tr}\left\{\rho u_{t} \dot{h}_t u_{t}^{\dagger}\right\}  \label{gauge-invariant work}
\end{equation}
and
\begin{equation}
Q_{inv}[\rho] = Q_{u}[\rho]  + Q_{c}[\rho]  
\label{gauge-invariant-heat}
\end{equation}
where the dot denotes the time derivative and
\begin{equation}
Q_{c}[\rho] = \int_{0}^{\tau} \dd t \operatorname{Tr}\left\{\rho\dot{u}_{t} h_t u_{t}^{\dagger}+\rho u_{t} h_t \dot{u}_{t}^{\dagger}\right\} 
\label{q-coherent}
\end{equation}
is the coherent heat. 
\end{theorem}
\indent These expressions are derived under the assumption that degeneracies remain constant over time. However, even when degeneracies vary over time, the expressions still hold. In this case, it is crucial to maintain consistency in the integration concerning the dimension of the thermodynamic group. To address this, we introduce the maximal gauge measure $\dd\mathcal{G}_{\operatorname{T}}^{\text{max}}$ (which corresponds to the scenario where all eigenvalues are degenerate). This measure can always be expressed as $\dd\mathcal{G}_{\operatorname{T}}^{\text{max}} = \dd \mathcal{G}_{\operatorname{T}}^c \dd\mathcal{G}_{\operatorname{T}}$, where $\dd\mathcal{G}_{\operatorname{T}}^c$ represents the complementary gauge measure relative to the maximal measure. Since the complementary measure is always normalized, integrating over the maximal volume reduces to the volume associated with the thermodynamic group. Consequently, by performing the integration over sub-intervals of $[0, \tau]$ where degeneracies remain unchanged, the results obtained in Ref.~\cite{gauge-lucas} are reproduced.

{Therefore, the existence of invariant quantities follow Definition~\ref{quantities}. In special, by the first law of Thermodynamics, we can obtain the relation between the coherent heat and the invariant work}
\begin{equation}
W_{inv}[\rho] = W_u[\rho] - Q_c[\rho].
\label{eq:work-coherence}
\end{equation} 
\indent \indent The gauge theory framework applied to the conventional concepts of heat and work divides the energy associated with work into two terms related to different physical phenomena. Quantum coherences, resulting from transitions in the energy eigenbasis, are fundamentally connected to irreversibility (heat generation). While this concept is not new, it naturally arises from the formalism as a direct consequence of the gauge group action.

As a final important comment on the hypothesis in~\ref{theo-lucas}, the results in Eqs.~\eqref{gauge-invariant work} and~\eqref{gauge-invariant-heat} are established under the condition that the unitary matrix $u_t$ is differentiable for all $t$. This condition ensures that the expressions for invariant work and heat do not change even if degeneracies vary over time. However, it is possible to develop an associated process when $u_t$ is not continuous. A particularly relevant physical case is the quench process, where the Hamiltonian suddenly changes at a given time. Quench dynamics is used to study a variety of physically interesting phenomena in the context of many-body physics, such as chaos~\cite{LMG-QPT1}, Anderson localisation~\cite{andesson}, many-body localisation~\cite{mb1,mb2}, and thermodynamic aspects of quantum phase transitions (QPTs)~\cite{gold-work, Campbell, Bento2024,Nascimento2024}. Following this, in the next section, we will study the invariant work and heat for quench processes and apply this to QPTs. 

{However, before this, let us see how the thermodynamic entropy emerges from our framework and also provide a physical interpretation of the invariant heat and work}. 

\subsection{\texorpdfstring{The $\mathcal{G}_{\operatorname{T}}$-entropy and second law}{gt-entropy and second law}}

\indent Remembering that our theory is built on the basis of Quantum Information Theory, it is natural to consider Von Neumann entropy as the measure of information, from which the invariant thermodynamic entropy emerges through Eq.~\eqref{invariant-2}. Formally, we  established the in the following.
\begin{theorem}
    ($\mathcal{G}_{\operatorname{T}}-$Entropy) The entropy that is invariant under the thermodynamic group is 
    \begin{equation}
        S_{\mathcal{G}_{\operatorname{T}}}[\rho(t)] = S_u[\rho_{dd}^{E}(t)], \label{gauge-entropy}
  \end{equation}
    where $\rho_{dd}^{E}(t)$ is defined in Eq.~\eqref{rhodd}.
\end{theorem}

The $\mathcal{G}_{\operatorname{T}}-$entropy is the von Neumann entropy for the average of the density matrix with respect to the thermodynamic group. This entropy is deeply connected with two quantum physical features: degeneracies and coherences. The connection with degeneracies is obvious, as it emerges from the definition of the thermodynamic group in Eq.~\eqref{isomorphism} and the invariant quantities in Eq.~\eqref{invariant-2}. On the other hand, the connection with coherences can be established by the relationship between $\mathcal{G}_{\operatorname{T}}-$entropy and the diagonal entropy, $S_d[\rho(t)] = S_u[\rho_{diag}]$, where $\rho_{diag}$ is the density operator with the off-diagonal elements (in the energy eigenbasis) removed.

Diagonal entropy was first defined in Ref.~\cite{Polkovnikov2008}, motivated by the fact that all information about any arbitrary average of observables is contained in the diagonal elements of the density matrix in the energy basis. This entropy satisfies several properties that are required for a thermodynamic entropy~\cite{Polkovnikov2008,gustavo}, and can be interpreted as quantifying the amount of randomness observed in the energy eigenbasis of the system. Its connection to the relative entropy of coherence $C[\rho]$ is given by~\cite{Chitambar2019}
\begin{equation}
C[\rho] = S_d[\rho] - S_u[\rho].
\end{equation}
\indent \indent In the particular case of a pure state, coherence is equal to the diagonal entropy since $S_u[\rho]=0$. However, the diagonal entropy is defined under the assumption that the system is non-degenerate, or that degeneracies are not relevant, as systems with degeneracy can lead to certain ambiguities in its definition~\cite{Polkovnikov2008}.

Note that $S_{\mathcal{G}_{\operatorname{T}}}$ contains only contributions from the diagonal elements of the density matrix in the energy eigenbasis. However, in the case where there are degeneracies, the elements associated with the group $\mathcal{U}(n_t^k > 1)$ contribute to the Haar average for $S_{\mathcal{G}_{\operatorname{T}}}$. Using this, and the expression Eq.~\eqref{gauge-entropy}, it is possible to write
\begin{equation}
   S_{\mathcal{G}_{\operatorname{T}}}[\rho(t)] =  S_{d}[\rho(t)]   +  S_{\Gamma}[f_t], \label{connections-entropy}
\end{equation}
where $S_{\Gamma}[f_t]  = -\operatorname{Tr}\{f_t\log(|f_t|)\}$ and $f_t$ are the block diagonal matrices
\begin{equation}
\begin{aligned}
      f_t = \begin{pmatrix}
        \begin{pmatrix}    
       \text{Non}\\\text{degenerate} \\ -\rho_{kk}(t)\\ 
        \end{pmatrix} & 0\\
        0 &  \begin{pmatrix}
        \text{Degenerate} \\  \dfrac{\operatorname{Tr}\{\rho^E_{n_t^k}(t)\}}{n_{k}} \\
        \end{pmatrix}
    \end{pmatrix}.
\end{aligned} \label{ft-express}
\end{equation}
\indent 
In this equation, $|f_t|$ means the absolute values of the elements of $f_t$.

We introduce $S_{\Gamma}[f_t]$ and $f_t$ just to simplify notation. In fact, $f_t$ is not a density matrix, and $S_{\Gamma}$ is not the von Neumann entropy. The relation in Eq.~\eqref{connections-entropy}, however, is highly helpful for studying the properties of $\mathcal{G}_{\operatorname{T}}-$entropy. Notably, $S_{\Gamma}[f_t]$ represents the difference between the contributions obtained by the Haar average and the usual contributions from the populations in the diagonal entropy. Specifically, if the energy spectrum is non-degenerate, the blocks of $f_t$ coincide, leading to $S_{\Gamma} = 0$. Consequently, the $\mathcal{G}_{\operatorname{T}}$-entropy reduces to the diagonal entropy. On the other hand, when the energy spectrum is completely degenerate, the first block of $f_t$ coincides with the diagonal density matrix in energy eigenbasis, and the relation in Eq.~\eqref{connections-entropy} reduces to $S_{\mathcal{G}_{\operatorname{T}}}$ in Eq.~\eqref{gauge-entropy}. 

The expression for $f_t$ in Eq.~\eqref{connections-entropy} is non-trivial as long as there is at least one non-degenerate energy level. However, even if the energy spectrum is completely degenerate, Eq.~\eqref{connections-entropy} is always valid, since $S_{\mathcal{G}_{\operatorname{T}}}\geq S_{diag}$, thus implying that $S_{\Gamma}\geq 0$. In this sense, we identify that the term $S_{\Gamma}$ precisely corresponds to the Holevo asymmetry measure~\cite{quantumt-twirling1, marvian} associated with the diagonal part of the density operator in the energy basis.

Moreover, we can prove that the equality in $S_{\mathcal{G}_{\operatorname{T}}} = S_{diag}$ is achieved if and only if the energy spectrum is non-degenerate. Consequently, the diagonal entropy naturally emerges from the $\mathcal{G}_{\operatorname{T}}$-entropy. From the point of view of the Holevo asymmetry measure, this implies that the states are completely symmetric~\cite{quantumt-twirling1} with respect to the thermodynamic group $\mathcal{G}_{\operatorname{T}}$.

Therefore, the relation between diagonal entropy and $S_{\mathcal{G}_{\operatorname{T}}}$ suggests interpreting $S_{\mathcal{G}_{\operatorname{T}}}$ as a generalized measure of the amount of randomness observed in the energy eigenbasis of both degenerate and non-degenerate systems. In this way, the Holevo asymmetry measure $S_{\Gamma}$ is associated with the amount of randomness introduced into the system due to degenerate energy levels, a pure quantum feature. Furthermore, we can see the similarity between the contributions of coherent heat in Eq.~\eqref{qc-split} and $\mathcal{G}_{\operatorname{T}}-$entropy. We have contributions associated with coherences and degeneracies, indicating a relationship between coherent heat and $\mathcal{G}_{\operatorname{T}}-$entropy. The proofs of all these results are provided in Appendix~\ref{g-entropy}.

In order to gain deeper insights into the physical aspects of our theory, we next apply it to two paradigmatic examples in many-body physics: the Landau-Zener and the Lipkin-Meshkov-Glick models.

\subsection{The physical meaning of $W_{inv}$ and $Q_{inv}$}

The theoretical foundation of our work is based on the gauge principle and its implications for the redefinition of thermodynamic quantities such as heat, work, and entropy. We start from the hypothesis that such quantities (as well as any other thermodynamic quantity) must emerge from a coarse-graining process, given that the density operator contains excessively detailed information for a thermodynamic description. This is akin to the classical case, where the microscopic state possesses too much information and the thermodynamic quantities emerge from the coarse-graining generated by the macroscopic measurements.

We define admissible transformations as unitary operations that preserve mean energy --- that is, those that commute with the Hamiltonian, forming what we call the thermodynamic group. This choice reflects the physical constraint that only measurements in the energy basis are accessible. Accordingly, we derive quantum thermodynamic quantities that are invariant under these transformations, which lie on an \emph{equilibrium} manifold in the sense of gauge transformations. 

We observe that this procedure can be generalized to other observables or even a set of observables. The structure of the group will change, implying changes in the thermodynamic quantities, but the theory is going to be the same. What is important here is that we are assuming that we do have a limited ability to control the system, which is always true from an experimental point of view when the system dimension increases too much (see, for instance, the experiment described in Ref.~\cite{Araujo2018} where quantum state tomography is simply impossible).

Such invariant quantities arise naturally in our formalism through Haar averages applied to functionals or to the density operator. The role of this average is to provide the coarse-graining from which the thermodynamic variables emerge. We therefore emphasize that the invariant quantities are not definitions, but direct consequences of the mathematical structure adopted. They emerge here in the same sense that the classical thermodynamic variables emerge from classical macroscopic measurements.

The case of an open system was discussed in Ref.~\cite{gauge-lucas} and leads to the usual definition of heat and work. Therefore, needs no extra discussion here. The case of deeper interest is the evolution of a closed quantum system and the idea of work and especially heat within this scenario. Therefore, we will concentrate on the case of a closed quantum system described by the possibly time-dependent Hamiltonian $H_t$ and by the density operator $\rho_t$. From the usual Liouville-von Neumann equation, the effective dynamics for the density operator $\rho_{\mathcal{G}}(t) \equiv \int \dd\mathcal{G}_{\operatorname{T}}V_t \rho(t) V_t^{\dagger}$ can be obtained via Haar measure, resulting in
\begin{equation}
    \dfrac{d\rho_\mathcal{G}(t)}{dt} = -i\hbar [H_t, \rho_\mathcal{G}(t)] + \mathcal{L}(\rho_t),
\end{equation}
with
\begin{equation}
  \mathcal{L}(\rho_t) = \int \dd\mathcal{G}_{\operatorname{T}} \left( \dot{V}_t \rho_t V_t^\dagger + V_t \rho_t \dot{V}_t^\dagger \right).  
\end{equation} 
The term $\mathcal{L}(\rho)$ encompasses the irreversibility induced by the \textit{coarse-graining}, thus reflecting our lack of ability to fully control the system. This term is associated with the time derivative of $V_t$ and, thus, is a consequence of the changes in the energy eigenbasis. Since we cannot measure all the observables, but just energy, the information used in the generation of coherences will be lost from our point of view, which we effectively interpret in terms of non-unitary dynamics. 

This effective dynamics makes clear the origin of the thermodynamic quantities appearing in our formalism
\begin{eqnarray}
\begin{cases}
    W_u[\rho_\mathcal{G}] = \displaystyle\int_0^{\tau} dt \operatorname{Tr}\left\{\rho_\mathcal{G}\dfrac{d H_t}{dt}\right\} = W_{inv}[\rho_t]\\ \\ 
    Q_u[\rho_{\mathcal{G}}] = \displaystyle\int_0^{\tau} dt \operatorname{Tr}\left\{H_t\dfrac{d \rho_\mathcal{G}(t)}{dt} \right\} = Q_c[\rho_t].
\end{cases} \label{thermo-quantities}
\end{eqnarray}
We see that $Q_c[\rho]$, the coherent heat, emerges from the non-unitary part $\mathcal{L}(\rho)$. This is not a consequence of the physical interaction between the system and some environment, but emerges from the coarse-graining. This makes information leak to places that we simply do not have access to. That is the origin of the irreversibility discussed here. As shown in Ref.~\cite{gauge-lucas}, $Q_c$ depends exclusively on the quantum coherences in the energy eigenbasis, implying that the restricted access to the energy eigenbasis induces an effective dynamics, which is non-unitary, where the cost of producing coherences is translated into energy lost, which we call heat. That is the effective cost of the coarse-graining. This is akin to the usual case of a system interacting with a heat bath, where energy (and information) flows to the bath, a place that we do not have access. 

Therefore, although we are dealing with a closed quantum system, the effective dynamics observed by any agent with limited access to it will be non-unitary and, thus, energy will be irreversibly lost. This is why this energy enters in our theory as heat.

The physical interpretation of the work is direct. It represents the accessible energy in the effective dynamics. As discussed in Ref.~\cite{gauge-lucas}, $W_{inv}$ is determined by the changes in the energy eigenvalues, showing that only these contribute to the work.

Let us discuss that our notions of work and heat are fully consistent with the ones presented in Ref.~\cite{Alicki1979}. Equation~\eqref{eq:work-coherence} establishes a connection with the first law of thermodynamics. $W_{inv}[\rho] = W_u[\rho] - Q_c[\rho]$ implies that the change in the internal energy can be written as $\Delta U = W_{inv}[\rho] + Q_c[\rho] + Q_u[\rho] =  W_{inv}[\rho] + Q_{inv}[\rho]$. This relation shows how the \textit{coarse-graining} splits the energy: part of the usual work ($W_u$) is converted into invariant work ($W_{inv}$) and part into coherent heat ($Q_c$). The sign of $Q_c$ determines the relation between $W_{inv}$ and $W_u$, indeed from Eq.~\eqref{eq:work-coherence}: 
\begin{align*}
Q_c > 0 &\Rightarrow W_{inv} < W_u \quad \text{(usual case with dissipation)} \\
Q_c < 0 &\Rightarrow W_{inv} > W_u \quad \text{(extra work extraction)}
\end{align*}

In the first case, energy is converted into irreversible contributions through \textit{coarse-graining}, reducing the accessible work, while in the second case, the system uses coherences in the energy basis to provide additional work. This is possible if the time dependence of the Hamiltonian is tied to some control parameter over which one has optimal control, allowing, in particular, for scenarios where techniques such as quantum control or specific constraints on the thermodynamic group make this feasible. This fact is important in small dimensions, where the level of control is high. However, this conversion of coherence into work decreases when the system becomes larger, and the level of controlling it decreases, vanishing in the thermodynamic limit, as it should. This is a purely quantum contribution to the heat.

Thus, based on these analyses, the physical meaning of the invariant work and heat, specially its coherent part, is clear and their connections to the well-established definitions in Ref.~\cite{Alicki1979} within the context of the first law of thermodynamics are established. 

Nevertheless, the term \emph{coherent heat} may sound strange since the evolution does not involve a real thermal bath, and the dissipation originates from the coarse-graining process via Haar averaging, manifesting through coherences in the energy basis and non-adiabatic changes of that basis. However, a more refined analysis shows that the term $\mathcal{L}(\rho)$ introduces effective dissipation, modifying the eigenvalues of $\rho_\mathcal{G}$ and generating entropy:
 \begin{equation}
        \dfrac{d}{dt}S_\mathcal{G}[\rho] = -\operatorname{Tr}\left\{\mathcal{L}(\rho)\log(\rho_\mathcal{G})\right\}
    \end{equation}
This suggests that $Q_c$ comes from the action of an effective heat bath. Moreover, $Q_c$ may behave like non-adiabatic work (especially when $Q_c < 0$), but such an interpretation neglects the underlying information loss. The terms adiabatic here mean slow changes, in such a way that no quantum coherence in the energy eigenbasis is produced; therefore, no transitions are induced into the system.

We conclude by stating that our theory splits the usual functional of heat defined by Alick into two contributions, one of which is associated with energy flux between the system and some physical environment and another one, associated with the generation of quantum coherences and information loss due to the coarse-graining introduced by the limited ability to measure the system. This last term is exactly what we call coherent heat.

An important aspect of our description concerns the limiting case in which our formulation is applied in the absence of coarse-graining. In this case, the thermodynamic group essentially reduces to a single element: the identity matrix corresponding to the system’s dimension. Here, the gauge principle acts trivially, such that any averaging over a given object, whether over functionals or operators, results in the object itself. Consequently, the invariant quantities in this limit coincide exactly with those presented in Ref.~\cite{Alicki1979} concerning work and heat.

\color{black}

\section{Thermodynamics of quantum phase transitions}
\label{sec:transitions}

Quantum phase transitions are rich phenomena exhibiting complex dynamics, making critical systems perfect for exploring Thermodynamics. We describe in general how our theory fits this scenario, developing novel concepts and discussing the physical meaning of our theory. In this context, we consider physical systems that can be described by the following Hamiltonian
\begin{equation}
    H_g = H_0 + g H_1. \label{hamil1}
\end{equation}
\indent \indent Here, $g$ is a controllable parameter and $H_0$ and $H_1$ are two Hermitian operators that generally do not commute. In particular, if these operators do not commute, the quantum fluctuations induced by the perturbation $H_1$ as $g$ changes may induce quantum phase transition (QPT) in the system at some critical value of $g$, which we denote as $g_c$~\cite{sachdev2011quantum}.

Usually, such transitions occur at the thermodynamic limit. However, signatures of critical behaviour for Hamiltonians of the type~\eqref{hamil1}, around the critical point $g_c$, can be observed in certain physical quantities, even in finite dimensions, particularly through dynamics following quantum quenches~\cite{Landi2021,pappalardi}. This dynamics generally involves preparing the system initially (at $t<0$) in the ground state $|\psi_0\rangle = |\psi(g_0)\rangle$ of the Hamiltonian $H_{g_0}$, associated with the initial parameter $g_0$. Then, the system is quenched to the final Hamiltonian $H_g$, where $g=g_0+\delta g$, with $\delta g$ denoting the quench amplitude, over a short time interval $\tau$ such that $\tau \to 0$~\cite{gold-work}. This implies that the initial state of the system remains the same, and only the Hamiltonian is modified.

In this sense, let us construct an analogous notion of a quench within the framework presented in the previous section. Indeed, we can construct a time-dependent Hamiltonian capable of emulating quench dynamics using a Heaviside function associated with the quench amplitude at $t=\tau$. Thus, we define the time-dependent version of Eq.~\eqref{hamil1} as follows
\begin{equation}
    H_{g}(t) = H_{g_0} + \delta g \theta(t-\tau) H_1 ,\label{TdH}
\end{equation}
where $\theta(t-\tau)$ is the Heaviside functions defined as $\theta(~t~-~\tau~)~=~1$ if $t\geq \tau$ and  $\theta(t-\tau) = 0$ if $t < \tau$. The system is prepared in ground state $\ket{\psi_0}$ of the Hamiltonian $H_{g_0}$. The Hamiltonian is them quenched, changing $g_0$ by $\delta g$. We are interested here in the work associated with the quench. From Hamiltonian Eq.~\eqref{TdH}, the invariant work associated with the quench reads 
\begin{equation}
        W_{inv}[\rho] = \dfrac{1}{2} \operatorname{Tr}\left\{ \rho^{E}_{diag} H_{g} - \rho^{E}_{dd} H_{g_0} \right\}, 
        \label{inv-work-qpt}
\end{equation}
in which we omit the dependence on $\tau$ in $\rho_{diag}^{E}, H_{g}$ and $\rho_{dd}^{E}$. The index $E$ in the density operators means that they are written in the $H_g$ eigenbasis.

Note that this expression is fundamentally distinct from the one obtained by employing Alicki's definition given in Eq.~\eqref{usual-work}. This is a fundamental consequence of the fact that we define work as a gauge invariant quantity, which forces coherences to be treated as part of the heat, thus leading to irreversibility.

Now, from Eq.~\eqref{inv-work-qpt} and the fact that the energy $U$ is also invariant, the first law of thermodynamics immediately leads us to the invariant heat
\begin{equation}
Q_{inv}[\rho] = Q_u[\rho] + Q_c[\rho],\label{qinv-qpt}
\end{equation}
where 
\begin{equation}
Q_{c}[\rho] = \dfrac{1}{2}\operatorname{Tr}\left\{\left(\rho^{E}_{dd}-\rho^{E} \right)H_{g_0}\right\}.
\label{inv-qc-qpt}
\end{equation}
\indent \indent The coherent heat given in Eq.~\eqref{inv-qc-qpt} is associated to two distinct aspects: coherences in energy eigenbasis and the degeneracies of the energy spectrum. This becomes clearer if we split the coherent heat as
\begin{equation}
    Q_c[\rho] = \dfrac{1}{2}\operatorname{Tr}\left\{ \left(\rho^{E}_{dd}-\rho_{diag}^{E} \right)H_{g_0}\right\}  -  \dfrac{1}{2}\operatorname{Tr}\left\{ \rho_{c}^{E} H_{g_0}\right\} \label{qc-split}
\end{equation}
 where $\rho^E_c$ is the density operator with the diagonal (in the energy eigenbasis) elements removed. In Appendix~\ref{Appendix-A}, we show that in the limit of non-degenerate energy levels, the first term of Eq.~\eqref{qc-split} vanishes, and all contributions arise from coherences in the energy basis, recovering the result obtained in Ref.~\cite{gauge-lucas}.

The gauge approach for quantum thermodynamics suggests that $Q_c[\rho]$ is indeed an energy term associated with heat. The expression for the invariant and coherent heat requires careful analysis for its physical interpretation. In particular, for closed systems, we can establish the mathematical connection between the average work (under the statistical notion of work) and the invariant work and heat as
\begin{equation}
 2Q_{c}[\rho] =   \langle W \rangle - 2W_{inv}[\rho]. \label{inner-friction}
\end{equation}
\indent \indent We see that the coherent heat is fundamentally connected to the notion of inner friction in quantum thermodynamics, a concept introduced to treat irreversibility in closed systems~\cite{inner}. Here, this quantity, which is associated with the changes in the energy eigenbasis, naturally emerges from the imposition of gauge invariance. 


\section{Applications}

In this section, we apply our formalism to two paradigmatic models. The first one, a non-degenerate two-level system, serves as a prototypical example for studying quantum phase transitions. The second model emerges in the context of many-body systems and features a phase characterised by degenerate energy levels.

\subsection{Landau-Zener}

Let us consider the dynamics of a single qubit described by the following Hamiltonian
\begin{equation}
    H_{LZ}(t) = \left(-\dfrac{\Delta}{2} + ag \right)\sigma^z + \epsilon \sigma^x. \label{Landau-Zener1}
\end{equation}
where $\sigma^{x}$ and $\sigma^{z}$ are Pauli matrices, and $g$ stands for the strength of the externally controlled magnetic field, whose coupling with the qubit is represented by $a>0$. $\epsilon$ is associated with the crossing ($\epsilon = 0$) or non-crossing ($\epsilon > 0$) energy levels and $\Delta$ is the bare frequency of the qubit.

\begin{figure}[ht!]
    \begin{center}
    \includegraphics[scale=0.5]{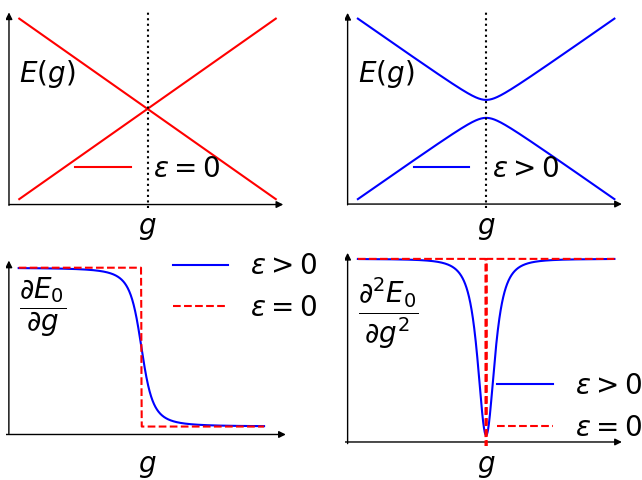}
    \caption{\justifying Schematic representation of the eigenenergies of the Landau-Zener Hamiltonian with and without energy crossing and the rate of variation of the ground state energy with respect to the control parameter $g$.}
    \label{general-LZ}
    \end{center}
\end{figure}

Hamiltonian Eq.~\eqref{Landau-Zener1} is the well-known Landau-Zener model, which is frequently employed in studies of phase transitions as it is a prototype for critical systems~\cite{quantum-quench, gold-work}. The system undergoes a quantum phase transition at the critical point $g_{c} = \Delta/2a$ and $\epsilon \to 0$. For $\epsilon > 0$, the transition is a first-order one, characterised by a discontinuous change in the order parameter. When $\epsilon = 0$, the transition becomes a second-order one, exhibiting a continuous change in the order parameter and associated critical fluctuations. A schematic representation of these situations is presented in Fig.~\ref{general-LZ}.

Based on the Hamiltonian~\eqref{TdH}, the time dependent version of Eq.~\eqref{Landau-Zener1} can be construct, resulting in
\begin{equation}
    H_{LZ}(t) = \left(-\dfrac{\Delta}{2} + a(g_0+\delta g \theta(t-\tau)) \right)\sigma^z + \epsilon \sigma^x. \label{Landau-Zener2}
\end{equation}
\indent \indent In this way, $H_{g_0} = \left(-\Delta/2 + ag_0\right)\sigma^z + \epsilon \sigma^x $ and  $H_1 =  a\sigma^z$. It is straightforward to compute both the invariant work and heat associated to this quench. In particular, since we are considering a closed system, the invariant heat reduces to the coherent heat. The work is given by 
\begin{equation}
    W_{inv}[\rho] = - \dfrac{a \delta g (a \delta g+\gamma_0) \left(a \delta g \gamma_0+\gamma_0^2+\epsilon ^2\right)}{2 \sqrt{\gamma_0^2+\epsilon ^2} \left((a \delta g+\gamma_0)^2+\epsilon ^2\right)}, \label{winv-explz}
\end{equation}
while the heat takes the form
\begin{equation}
    Q_c[\rho] =  \dfrac{a^2 \delta g^2 \epsilon ^2}{2 \sqrt{\gamma_0^2+\epsilon ^2} \left((a \delta g+\gamma_0)^2+\epsilon ^2\right)},\label{qc-explz}
\end{equation}
where $\gamma_0 \equiv ag_0 -\Delta/2$. We provide the details of these calculations in Appendix~\ref{Appendix-B1}.
\begin{figure}[ht!]
    \centering
   \includegraphics[scale=0.5]{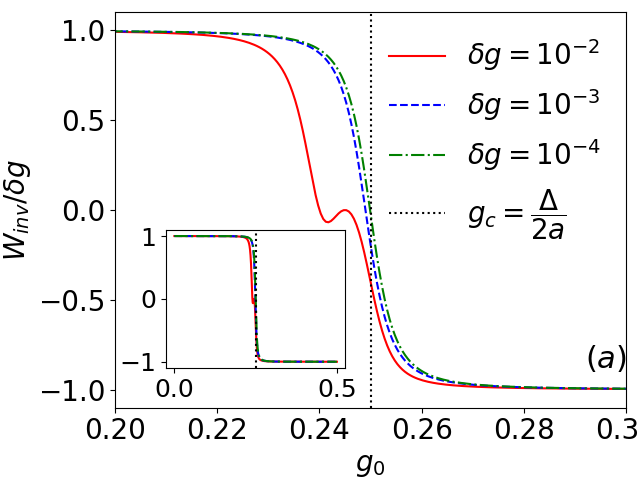}\\
   \includegraphics[scale=0.5]{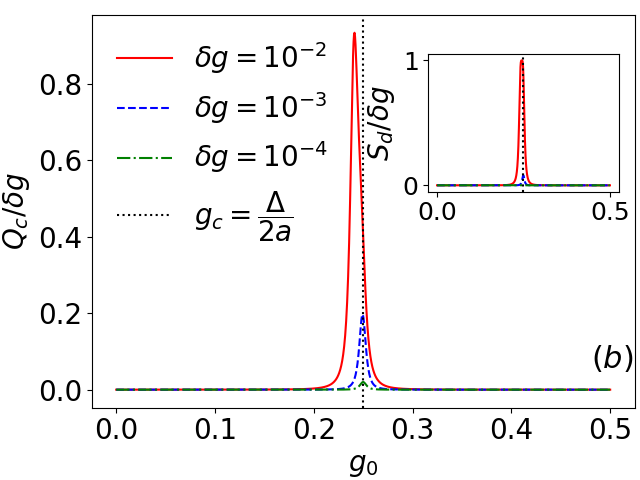}
  \caption{\justifying Invariant work (a), heat (b) and diagonal entropy per quench for Landau-Zener model. In the inset of coherent heat, we plot the diagonal entropy per quench. We set $a=2$, $\Delta =1$ and  $\epsilon = 0.001$ for different values of the quench amplitude $\delta g$. The dotted vertical lines marks the critical point of the model.}
    \label{lz-complete}
\end{figure}

Figure~\ref{lz-complete} reveals the behaviour of the work [panel~\ref{lz-complete}(a)], heat [panel~\ref{lz-complete}(b)], and entropy [panel~\ref{lz-complete}(c)] for distinct quenches from $g_0$ to $g = g_0 + \delta g$. In particular, panel~\ref{lz-complete}(a) shows that the invariant work exhibits a similar behaviour to that observed for the derivative of the ground state energy of the model, as shown in Fig.\ref{general-LZ}. This similarity is especially pronounced in the regime of small quenches, i.e., $\delta g \to 0$. Furthermore, for small quenches, panel~\ref{lz-complete}(b) indicates that the coherent heat approaches zero, suggesting a negligible contribution of energy associated with the change of basis during the quench. Indeed, for $Q_c[\rho] \approx 0$ and considering the system with no degeneracies, we can connect $W_{\text{inv}}[\rho]$ with the variation of the ground state energy with respect to the parameter $g_0$ as
\begin{equation}
\dfrac{dE_0}{dg_0} \approx \dfrac{2}{g_0} W_{\text{inv}}[\rho], \label{hellman-feynman}
\end{equation}
which follows from the Hellman-Feynman Theorem~\cite{Feynman}. The expression in Eq.~\eqref{hellman-feynman} justifies the correspondence between the variation of the ground state energy and the invariant work in Figs.~\ref{general-LZ} and~\ref{lz-complete}(a).

Another interesting aspect to consider in the Landau-Zener model is the case with energy level crossing, i.e., $\epsilon = 0$. In this case, the contribution of heat is zero, which is expected since $[H_0, H_{\tau}] = 0$. However, the model exhibits degeneracy at the critical point $g_0 = \Delta/4a$. At this point, the thermodynamic group $\mathcal{G}_{\operatorname{T}}$ is isomorphic to the group $\mathcal{U}(2)$. Consequently, the Haar measure induced by the thermodynamic group is $\dd\mathcal{G}_{\operatorname{T}}^{g_c} = \dd\mu\left[\mathcal{U}(2)\right]$. Therefore, it follows that the expression for the invariant work is modified to $W_{\text{inv}}[\rho] = - a\delta g/2 \operatorname{sgn}\left(g_0 - \Delta/2\right)$, where $\operatorname{sgn}(\cdot)$ denotes the sign function.

Signatures of the QPT can be seen in the derivatives of invariant work and heat. As the system approaches the critical point of the phase transition, these quantities exhibit divergent behaviour. This divergence is visually evident in Fig.~\ref{dlz-complete}. These expressions are all obtained in Appendix~\ref{Appendix-B}.

\begin{figure}[ht!]
    \centering
    \includegraphics[scale=0.5]{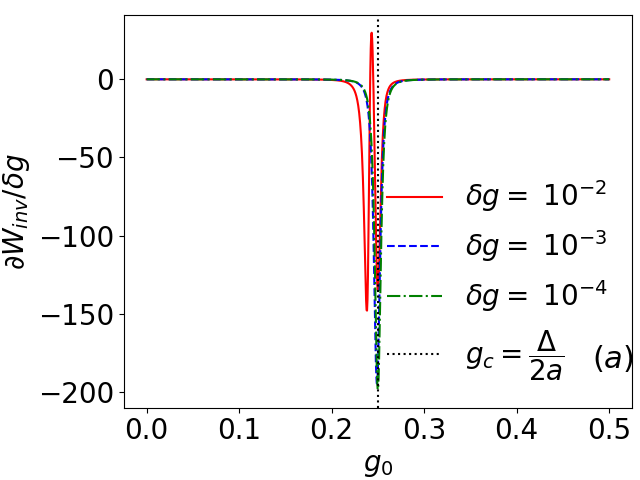}
    \includegraphics[scale=0.5]{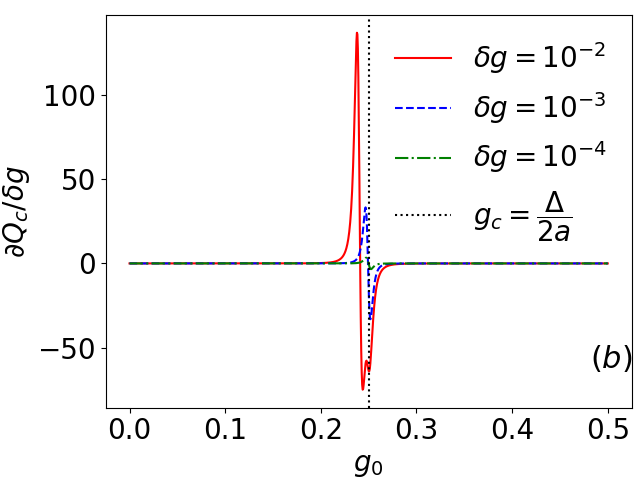}
    \caption{\justifying Derivative of invariant work (a) and heat (b) for different quenches for the Landau-Zener model. In these plots, we set $a = 2$, $\Delta = 1$, and $\epsilon = 0.001$. The divergence appearing in the invariant work (coherent heat) for small (large) values of the  quench amplitude is evident.}
    \label{dlz-complete}
\end{figure}

\subsection{Lipkin-Meshkov-Glick (LMG)}

\indent Let us now consider a more complex system, exhibiting degenerate eigenlevels and quantum phase transition, described by the Hamiltonian
\begin{equation}
    H_g = -\dfrac{k}{2j}\left(J_z^2 + \gamma J_y^2\right) -gJ_x ,\label{lmg-1}
\end{equation}
where $J_{\alpha} = \sum_{i=1}^{N} \sigma^{\alpha}_{i}/2;$ $(\alpha = x,y,z)$ are the collective spin operators, with $\sigma_{i}^{\alpha}$ denoting the $\alpha$ Pauli matrix acting on the $i$-th site of a chain of $N = 2j$ sites, and $j$ is the total angular momentum. The parameters $k > 0$, $0 \leq \gamma \leq 1$, and $g \geq 0$ denote the constants associated with the coupling between the spins, the anisotropy of the model, and the strength of the magnetic field along the $x$-axis, respectively.

The Hamiltonian in Eq.~\eqref{lmg-1} is a particular instance of the well-known Lipkin-Meshkov-Glick model~\cite{lmg1,lmg2,lmg3}. {In this work we consider only the fully symmetric subspace of this model, thus reducing the dimension from $2^{N}$ to $N+1$}. Various properties of the LMG model have already been studied in the literature in different contexts such as nuclear physics~\cite{lmg-nuclear}, optics~\cite{lmg-optica}, quantum information~\cite{lmg-spread}, and condensed matter~\cite{lmg-condensed}, just to mention a few. Therefore, given its wide applicability, it is the right model for us to consider in terms of its quantum thermodynamics in the context of gauge invariance.

\begin{figure}[ht!]
    \centering
    \includegraphics[scale=0.53]{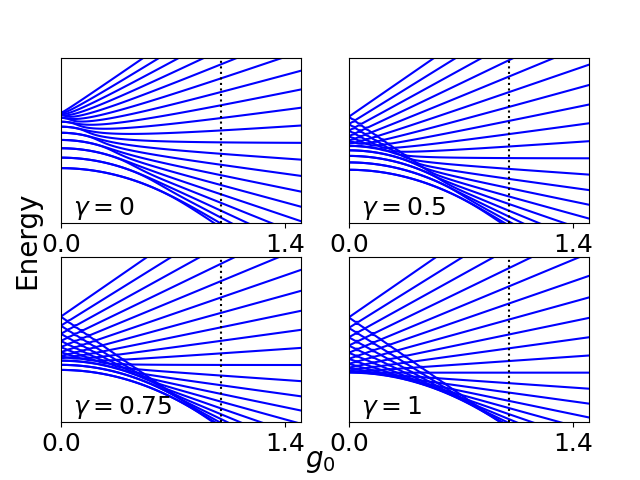}
    \caption{\justifying Energy spectrum of LMG model in Eq.~\eqref{lmg-1} for different values of $\gamma$. Here we used $j=10$ for all panels. The vertical dotted line marks the critical point of the model.}
    \label{lmg-spectrum}
\end{figure}

In the thermodynamic limit, i. e., $j \to \infty$, the model presents a quantum phase transition~\cite{pappalardi, Campbell} where the parity symmetry is broken as the magnetic field changes from $g < g_c$ (ferromagnetic phase) to $g > g_c$ (paramagnetic phase), where $g_c = k$ is the critical point of the equilibrium phase transition \cite{Campbell}. A particular feature of these two phases is related to degeneracies of the Hamiltonian. While the ferromagnetic phase is doubly degenerate, the paramagnetic phase presents no degeneracy. Figure.~\ref{lmg-spectrum} shows such configurations. 


Given this feature, the invariant work, heat and entropy does not reduce to the simple form that we saw in the Landau-Zener model. The average of the density matrix over the thermodynamic group assumes the general form given in Eq.~\eqref{rhodd}. In order to apply our theory to the case in hand, we consider the time dependent counterpart associated to the Hamiltonian Eq.~\eqref{lmg-1} as

\begin{equation}
H_{g}(t) = -\dfrac{k}{2j}\left(J_z^2 + \gamma J_y^2\right) - (g_0+\delta g \theta(t-\tau))J_x, \label{LMGh}
\end{equation}

\indent \indent We then have $H_{g_0} = -(k/2j)\left(J_z^2 + \gamma J_y^2\right) - g_0J_x$ and $H_1  = -J_x$. In the present case, it is not possible to obtain analytical expressions for the eigenvectors and/or eigenvalues in terms of the control parameter $g$ as we did for the Landau-Zener model. Therefore, we investigate this model using numerical methods. However, before proceeding, we will rewrite the expressions for the invariant work and heat in the more convenient form (See appendix \ref{Appendix-A})
\begin{equation}
W_{inv}[\rho] = -\dfrac{\delta g}{2} \operatorname{Tr}\left\{\rho_{dd}^{E} J_x \right\} \label{inv-work-lmg}
\end{equation}
and
\begin{equation}
Q_{inv}[\rho] = -\dfrac{\delta g}{2} \operatorname{Tr}\left\{\left(\rho^E - \rho_{dd}^{E} \right)J_x \right\} .\label{inv-heat-lmg} 
\end{equation}
\indent \indent Here, we proceed by fixing the quench amplitude at $\delta g = 0.01$ and the anisotropy constant at $\gamma = 0.75$. Previous numerical analyses have shown that the qualitative behavior of the quantities is independent of these choices.

\begin{figure}[ht!]
    \centering
    \includegraphics[scale=0.5]{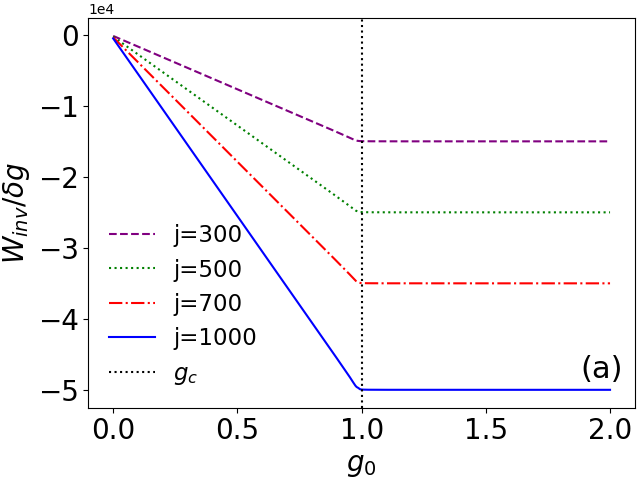}
    \includegraphics[scale=0.5]{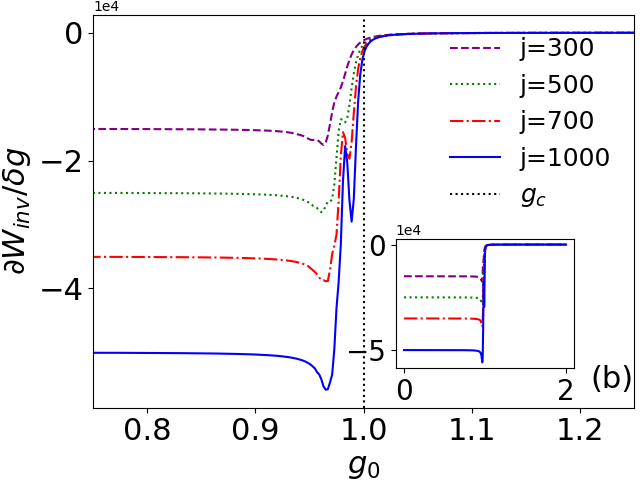}
    \includegraphics[scale=0.5]{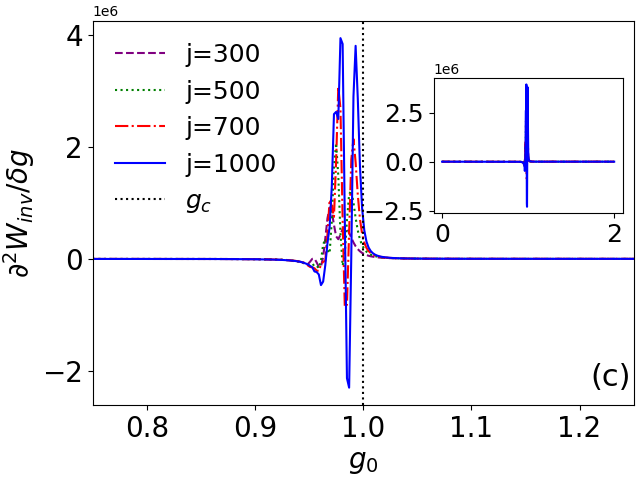}
    \caption{\justifying Thermodynamics of the LMG model. Panel (a) shows the invariant work, normalised by the quench, considering distinct values of $j$. Next, we see in Panel (b) the first derivative of the invariant work with respect $g_0$, normalised by quench amplitude. Panel (c) revels the divergence of second derivative of the invariant work in vicinity of  the quantum critical point.}
    \label{fig:inv-work-lmg}
\end{figure}

A direct consequence of the theory emerges from the expression for the invariant work. It is an order parameter for the system, since the magnetisation $\langle J_\alpha \rangle = \operatorname{Tr}\{\rho J_{\alpha}\}$ (with $\alpha=x,y,z$) is the usual order parameter signalling the quantum phase transition in the this case~\cite{gold-work}. We show such behaviour in Fig.~\ref{fig:inv-work-lmg}(a), where the order parameter feature of the work is clear. The other panels in this figure show the dynamical behaviour of the first and second derivatives of the work with respect to the quench parameter $g_0$. While the second derivative diverges near the critical point, the first derivative exhibits oscillations, which are a consequence of the changes in the degeneracies as we change $g_0$. It is clear from these figures that the invariant work is deeply linked to the quantum phase transition.

\begin{figure}[ht!]
    \centering
    \includegraphics[scale=0.5]{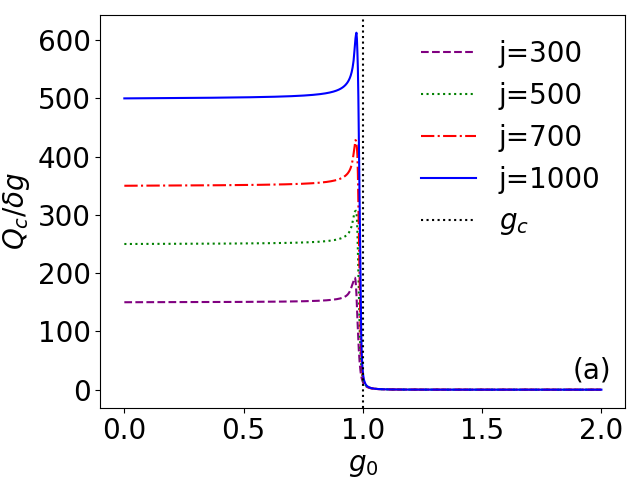}
    \includegraphics[scale=0.5]{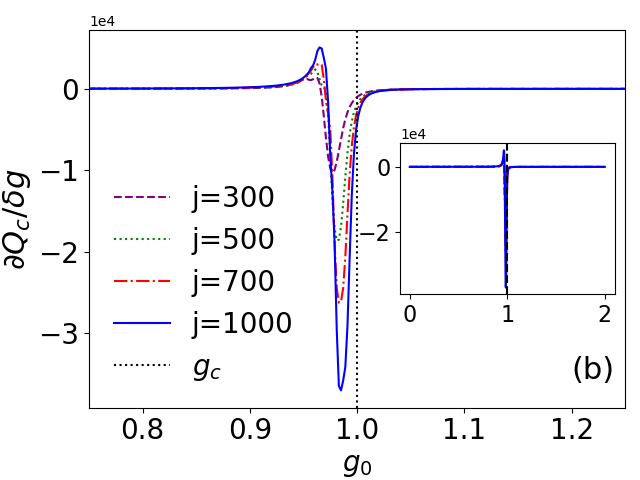}
    \caption{\justifying Invariant heat in Panel (a) and its derivative in Panel (b), for the LMG model considering different values of $j$ per quench. }
    \label{fig:inv-heat-lmg}
\end{figure}

Another interesting result that follows form the theory is associated to the invariant heat, which is shown in Fig.~\ref{fig:inv-heat-lmg}. We can see that the heat suffers and abrupt change near the critical point of the model, while the divergence of its derivative with respect to the quench clearly signs this point. 

\begin{figure}[!ht]
    \centering
    \includegraphics[scale=0.5]{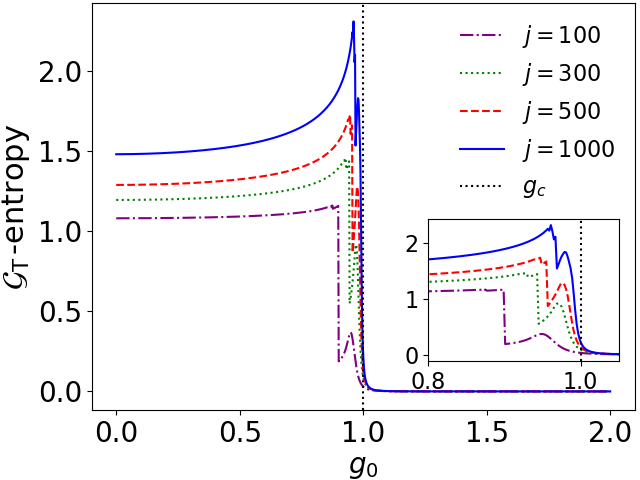}
    \caption{\justifying Thermodynamic $\mathcal{G}_{\operatorname{T}}-$entropy for the ground state of the LMG model as function of $g_0$ considering distinct values of $j$. The inset plot shows a zoom in the vicinity of the critical point, highlighting the abrupt change in the $S_{\mathcal{G}_{\operatorname{T}}}$ at criticality.}
    \label{fig:all-entropy}
\end{figure}

Now we discuss the thermodynamic entropy $S_{\mathcal{G}_{\operatorname{T}}}$ for the case at hand, which is not equivalent to the diagonal entropy as in the Landau-Zener model, since we have degeneracies. As mentioned earlier, the ferromagnetic phase ($g < g_c$) is doubly degenerate, while the paramagnetic phase ($g > g_c$) is completely non-degenerate. Therefore, we expect very distinct behaviours of $S_{\mathcal{G}_{\operatorname{T}}}$ in both phases. This is shown in Fig.~\ref{fig:all-entropy}. 

We first observe the similarity between the $\mathcal{G}_{\operatorname{T}}$-entropy in Fig.~\ref{fig:all-entropy} and the heat displayed in Fig.~\ref{fig:inv-heat-lmg}(a). This fact is not a coincidence but a consequence of the theory that makes both quantities directly associated with the production of quantum coherence in the energy eigenbasis, as well as with the presence of degeneracies in the ferromagnetic phase. In particular, this result confirms and extends the understanding of heat proposed in Ref.~\cite{gauge-lucas}.

Moreover, Fig.~\ref{fig:all-entropy} clearly shows a tendency for an abrupt change in $S_{\mathcal{G}_{\operatorname{T}}}$ in the vicinity of the critical point. Such change occurs precisely at the point where the degeneracies of the system are completely extinguished due to the competition between the contributions of the spin coupling and the external magnetic field. The extinction of degeneracies before the critical point occurs because we are dealing with a finite-dimensional system. 

However, as shown in the Fig.~\ref{fig:all-entropy}, this point tends to shift to the critical point of the model for larger dimensions, where we approach the thermodynamic limit. Additionally, we observe that the abrupt change in $S_{\mathcal{G}_{\operatorname{T}}}$ is more pronounced in chains with fewer spins, which is associated with the contributions of degeneracies being more significant in systems further away from the thermodynamic limit.

\begin{figure}[ht!]
    \centering
    \includegraphics[scale=0.5]{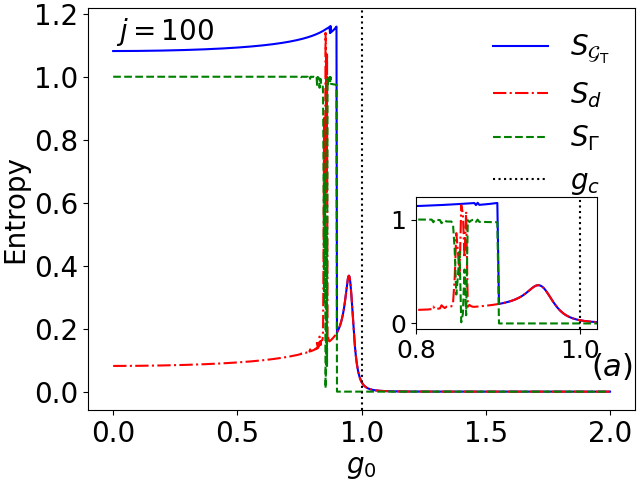}
    \includegraphics[scale=0.5]{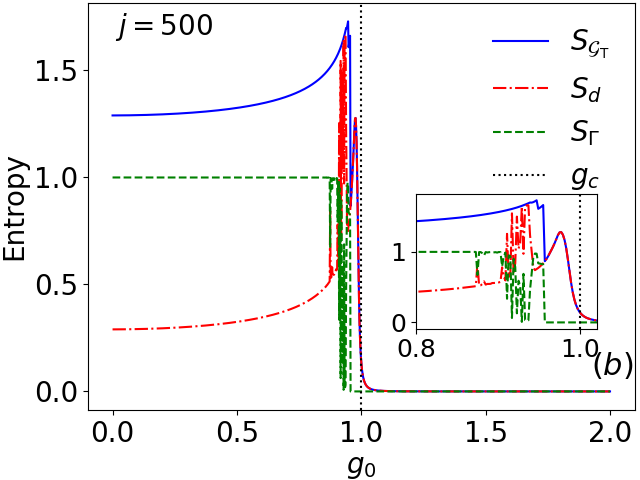}
    \includegraphics[scale=0.5]{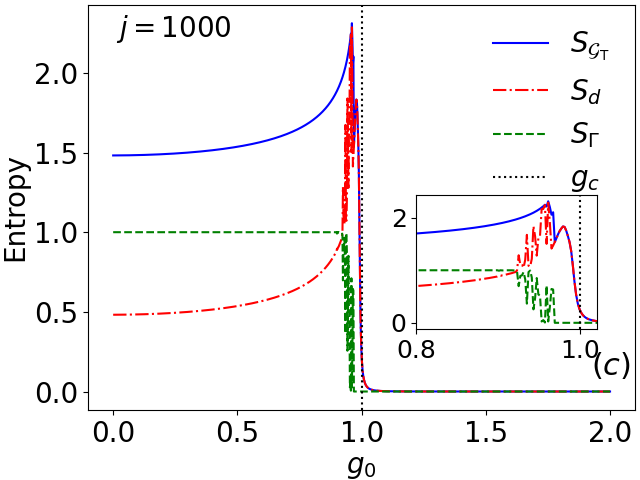}
    \caption{\justifying Comparative of thermodynamic gauge entropy $S_{\mathcal{G}_{\operatorname{T}}}$, diagonal entropy $S_{d}$ and Holevo asymmetry measure $S_{\Gamma}$ for $j=100, 500, 1000$ in Panels (a), (b) and (c) respectively. The insets in all panels correspond to a zoom in the vicinity of the critical point.}
    \label{fig:gauge-entropy}
\end{figure}

Accordingly, we analyze $S_{\mathcal{G}_{\operatorname{T}}}$ in comparison with the diagonal entropy $S_d$, highlighting the contributions of degeneracies as quantified by the Holevo asymmetry measure $S_{\Gamma}$ and also the effect of the size of the chain. First, we observe from Fig.~\ref{fig:gauge-entropy} that oscillatory behavior of the Holevo asymmetry measure in the vicinity of the critical point arises from the breaking of energy level degeneracies, which results in modifications to the structure of the thermodynamic group $\mathcal{G}_{\operatorname{T}}$. As the value of $j$ increases, the oscillations in the Holevo asymmetry $S_{\Gamma}$ decrease near the critical point and eventually vanish at the quantum phase transition. 

\begin{figure}[ht!]
    \centering
    \includegraphics[scale=0.5]{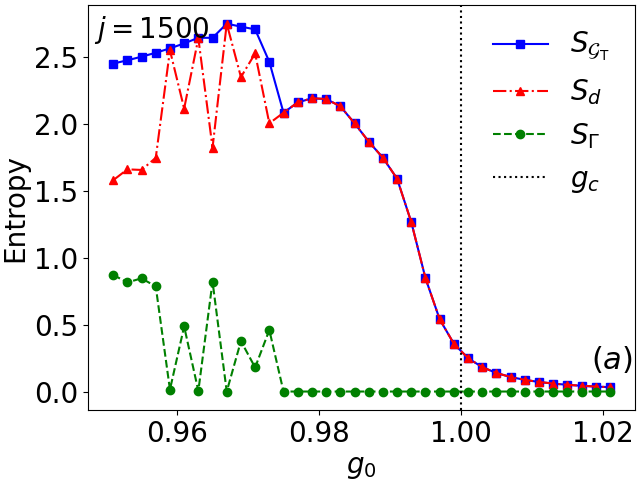}
    \includegraphics[scale=0.5]{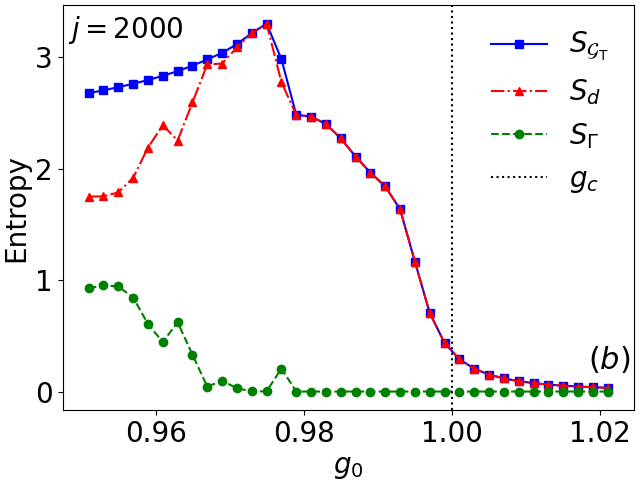}
    \includegraphics[scale=0.5]{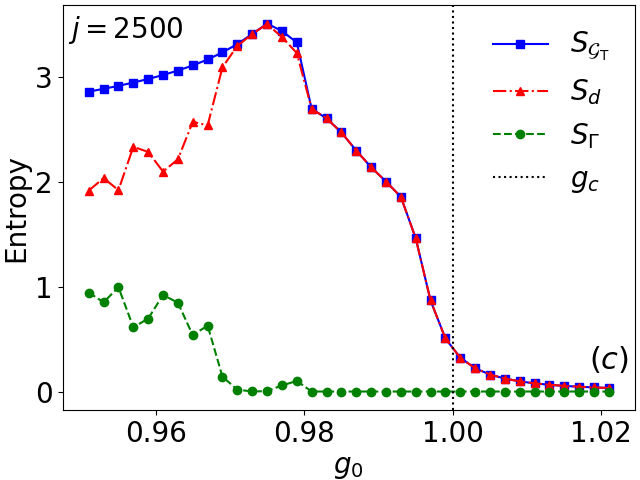}
    \caption{\justifying Comparative of Thermodynamic gauge entropy $S_{\mathcal{G}_{\operatorname{T}}}$, diagonal entropy $S_{d}$ and Holevo asymmetry measure $S_{\Gamma}$ in the vinicity of the quantum critical point in the LMG model for $j=1500, 2000, 2500$ in Panels (a), (b) and (c) respectively.}
    \label{fig:vicinity-LMG}
\end{figure}

Then, from the point of view of the Holevo asymmetry measure, degeneracies produce asymmetry and consequently increase the randomness and entropy of the state. However, it is the relationship between the total volume of the subspaces $\mathcal{H}_k$ and the changes in degeneracies that leads to abrupt modifications in the thermodynamic entropy. Consequently, we expect a greater impact of degeneracies in lower-dimensional systems, since the Haar averages contained in $S_{\Gamma}$ are significant with respect to the contributions of populations coming from the diagonal entropy.
\indent  However, this scenario is modified as we consider systems with larger dimensions, as already indicated in Fig.~\ref{fig:gauge-entropy}(c). In this regard, in Fig.~\ref{fig:vicinity-LMG} we study how the entropies behave in the vicinity of the critical point when we consider higher dimensional models.
 Thus, for large values of $j$, near the critical point of the phase transition, the diagonal entropy approaches $S_{\mathcal{G}_{\operatorname{T}}}$, while the Holevo asymmetry fluctuates around values close to zero. Indeed, this behaviour emerges because the contributions of the remaining degeneracies are not as significant as in smaller dimensions, as shown by the $S_{\Gamma}$ curve in Figs.~\ref{fig:gauge-entropy} and~\ref{fig:vicinity-LMG}. Therefore, from Eq.~\eqref{connections-entropy} it follows that near the critical point $|S_{\mathcal{G}_{\operatorname{T}}} - S_{d}| = S_{\Gamma}$ and, since $S_{\Gamma}$ becomes increasingly less significant, we conclude that the $\mathcal{G}_{\operatorname{T}}$-entropy and the diagonal entropy approach each other. Thus, the Holevo asymmetry acts as a parameter that determines the continuity of the thermodynamic gauge entropy.
 
On the other hand, for quenches occurring after the critical point of the phase transition, we observe that $S_{\mathcal{G}_{\operatorname{T}}}$ asymptotically tends to zero. Indeed, after the critical point, the thermodynamic group is always given by $\mathcal{G}_{\operatorname{T}} = \prod_{k=1}^{d}\mathcal{U}(1)$, and consequently, the diagonal part of the states $\rho$ in the energy basis becomes increasingly symmetric with respect to the group $\mathcal{G}_{\operatorname{T}}$, thereby asymptotically reducing the entropy.


\section{Discussion}
\label{sec:discussions}

In this article, we develop a theory of Quantum Thermodynamics as a gauge theory of the thermodynamic group, following the approach outlined in~\cite{gauge-lucas}. This method is rooted in the gauge invariance principle, where thermodynamic quantities arise within the context of Haar averages. As a result, invariant forms of heat, work, and entropy are defined for general quantum systems. This provides a comprehensive theory of Quantum Thermodynamics grounded in the fundamental principle of gauge invariance. To demonstrate some physical implications of this framework, the theory is applied to quantum critical systems.

A key result of the theory is the unique decomposition of a system's energy into work and heat, addressing one of the most debated issues in Quantum Thermodynamics. Specifically, work is inherently linked to populations, while heat emerges from the creation of quantum coherences in the energy eigenbasis. This allows heat to be interpreted as a form of internal friction~\cite{inner}, a concept largely used in literature to justify irreversibility in closed quantum systems. Moreover, both heat and work are directly influenced by the degeneracy structure of the energy spectrum. As a result, the theory naturally incorporates distinct quantum characteristics in defining thermodynamic quantities.

Another fundamental result is the formulation of thermodynamic entropy. By interpreting von Neumann entropy as a measure of information in Quantum Mechanics, a single thermodynamic entropy emerges, which remains invariant under the action of the thermodynamic group. This entropy can be divided into two components: the first is the well-known diagonal entropy~\cite{Polkovnikov2008}, while the second is a purely quantum contribution resulting from the system's degeneracies. The gauge invariant entropy derived from this theory was shown to satisfy the essential properties required for thermodynamic entropy. Notably, in the classical limit, this last quantum contribution disappears, reducing the gauge invariant entropy to the diagonal entropy. Consequently, the diagonal entropy naturally arises from our formalism when considering non-degenerate, or classical Hamiltonians.

Additionally, thermodynamic entropy is fundamentally connected to the generation of quantum coherences in the energy eigenbasis, which naturally corresponds with invariant heat. It is important to note that these results were not predetermined but arose naturally from the theory. This reinforces the physical consistency of the theory, as Thermodynamics traditionally associates heat with entropy generation.

The relationship between heat and quantum coherences is neither novel nor unexpected~\cite{Polkovnikov2011}. From a thermodynamic point of view, our ability to control a system is in general limited. For instance, in many experiments, we are often restricted to measuring only a few observables, typically energy. Consequently, when quantum coherences are produced, the uncertainties in these observables increase, leading to entropy generation. Within the gauge framework presented here and in Ref.~\cite{gauge-lucas}, this connection arises naturally.

All of these aspects become evident when we apply the theory to well-known quantum critical systems: the Landau-Zener and LMG models. While the Landau-Zener model describes a non-degenerate two-level system, the LMG Hamiltonian represents fully connected spin chains with a doubly degenerate phase. In all cases, the invariant thermodynamic quantities defined here proved to be highly sensitive to the quantum phase transitions in these models. Moreover, the quantum nature of thermodynamics becomes clear.

{A natural question arising from our work regards a possible Clausius-like inequality associated with $Q_{inv}[\rho]$. Preliminary analysis suggests such a relation indeed exists. The natural path to approach this problem seems to involve the gauge entropy introduced in this work as a fundamental quantity. In this approach, the heat flux would have two distinct components: $J_u$, associated with exchange with real thermal baths, and $J_{\mathcal{G}}$, corresponding to the effective flux induced by the coarse-graining process. The resulting generalized version is expected to reduce to the conventional Clausius inequality in cases where $Q_c$ behaves like traditional heat, while adequately capturing the irreversible effects associated with informational constraints. The full details of this analysis will be the subject of future investigation.}

An intriguing question arises regarding the observed connection between $S_{\Gamma}$, the contribution of degeneracy to entropy, and the Holevo asymmetry measure. In the model under consideration, this connection is linked to changes in the symmetry of the Hamiltonian. This prompts us to ask whether such a connection is also present in other critical systems, such as those exhibiting dynamical criticality~\cite{Heyl2018}. Further investigation in this direction could deepen our understanding of Thermodynamics in the quantum regime.

Additionally, applying this framework to other contexts, such as heat engines, could provide new insights into the design and operation of these devices. For example, the theory clearly indicates that the efficiency of a heat engine is directly linked to the processes employed. This is because entropy and heat are associated not only with the non-unitary parts of the engine but also with the unitary ones. If the non-unitary processes generate coherences ---often due to being performed in a non-adiabatic way--- the efficiency of the engine will decrease.

\begin{acknowledgments}
We thank Philipp Strasberg for stimulating discussions and for critically reading the manuscript. This work was supported by the National Institute for the Science and Technology of Quantum Information (INCT-IQ), Grant No. 465469/2014-0, by the National Council for Scientific and Technological Development (CNPq), Grants No 308065/2022-0, and by Coordination of Superior Level Staff Improvement (CAPES). \L R acknowledges IRA Programme, project no. FENG.02.01-IP.05-0006/23, financed by the FENG program 2021-2027, Priority FENG.02, Measure FENG.02.01., with the support of the FNP.
\end{acknowledgments}

\appendix

\section{Haar measure and invariant thermodynamics quantities}
\label{Appendix-A}

\indent In this appendix, we will present some results associated with Haar measure theory~\cite{nachbin}. Subsequently, we will show the general calculation for determining the density operator $\rho_{dd}$ defined in Eq.~\eqref{rhodd}.  Finally, we will derive Eqs.~\eqref{inv-work-qpt} and~\eqref{qinv-qpt}.

\begin{definition}
(Haar \cite{haar-information}). The Haar measure on the unitary group $\mathcal{U}(d)$ is the unique normalized probability measure $\dd\mu \equiv \dd\mu[\mathcal{U}(d)]$ that is both left and right invariant over the group $\mathcal{U}(d)$, i.e., for all integrable functions $f$ and for all $V \in \mathcal{U}(d)$, we have:
\begin{align}
    \int_{\mathcal{U}(d)}\dd\mu \; f(\mathcal{U})&=\int_{\mathcal{U}(d)}\dd\mu f(\mathcal{U} V) \label{e0} \\
    &=\int_{\mathcal{U}(d)} \dd\mu \; f(V \mathcal{U}). \label{e00}
\end{align}
\end{definition}
In the particular case of $f=\rho$, where $\rho \in \mathcal{H}^d$ is arbitrary, we can evaluate the Haar integral of $\rho$ over the unitary group $\mathcal{U}(d)$ explicitly. Let us consider the linear transformation $D(\rho)$ defined by
\begin{equation}
    D(\rho) = \int_{\mathcal{U}(d)}  \dd\mu\;  \mathcal{U} \rho \mathcal{U}^{\dagger}. \label{e1}
\end{equation}
To do so, let us consider $ V \in \mathcal{U}(d) $ arbitrary, then, using left and right invariance of Haar measure we can write
\begin{align}
V D(\rho) V^{\dagger} &= V\left[\int_{\mathcal{U}(d)}\dd\mu\; \mathcal{U} \rho \mathcal{U}^{\dagger} \right] V^{\dagger}  \nonumber\\
&= \int_{\mathcal{U}(d)}\dd\mu \; (V \mathcal{U}) \rho(V \mathcal{U})^{\dagger} \nonumber\\
&= \int_{\mathcal{U}(d)}\dd\mu \; \mathcal{U} \rho \mathcal{U}^{\dagger}= D(\rho). \label{e2}
\end{align}
\indent  Given that $V D(\rho) V^{\dagger} = D(\rho)$ implies $[D(\rho),V] = 0$ for all $V \in \mathcal{U}(d)$, the Schur orthogonality relations \cite{sanmartin-grupodelie} establish $D(\rho) = C\mathbb{1}_d$, where $C$ is a proportionality constant. Since $ D(\rho) $ is diagonal, it can be characterized it by its trace, thus
\begin{align}
\operatorname{Tr}\{D(\rho)\} &=\operatorname{Tr}\left\{\int_{\mathcal{U}(d)} \dd\mu \; \mathcal{U} \rho \mathcal{U}^{\dagger}\right\}  \nonumber \\
&=\int_{\mathcal{U}(d)}\dd\mu \; \operatorname{Tr}\left\{\mathcal{U}\rho \mathcal{U}^{\dagger}\right\}  = \operatorname{Tr}\{\rho\}. \label{e3}
\end{align}
\indent  Therefore, Eq.~\eqref{e3} leads to $C =\operatorname{Tr} \{ \rho \}/d $, consequently, Eq.~\eqref{e1} takes the form
\begin{equation}
D(\rho) = \int_{\mathcal{U}(d)}  \dd\mu \; \mathcal{U} \rho \mathcal{U}^{\dagger}  = \dfrac{\operatorname{Tr} \{ \rho \}}{d}\mathbb{1}_{d}, \label{ef}
\end{equation}
for any arbitrary matrix $\rho$.  

When the Haar measure is induced by a group as in Eq.~\eqref{isomorphism}, it decomposes as a product of measures $\dd\mu_k \equiv \dd\mu[\mathcal{U}(n_t^k)]$ associated with each subspace $\mathcal{H}_k \in \mathcal{H}^d$, and the notion of the integral is extended by Fubini's theorem \cite{Fubbini-Halmos}, resulting in a multidimensional integral. Now, let us evaluate the Haar average of $\rho \in \mathcal{H}^d$, in the energy basis of $H_t$, and $\mathcal{H}^d = \bigoplus_{k=1}^{p} \mathcal{H}_k$ where $p\leq d$, over the thermodynamic group, whose Haar measure is given in Eq.~\eqref{haar-measure} and the unitary transformations are given in Eq.~\eqref{unitary-gauge-transformation}. Let $\mathcal{I}_k$ be the trivial embedding of $\mathcal{H}_k$ into $\mathcal{H}^d$, then it follows that
\begin{align}
    \Lambda_{\mathcal{G}_{\operatorname{T}}}[\rho^E]    &=  \int \dd\mathcal{G}_{\operatorname{T}} \mathcal{V}_{t} \rho^E \mathcal{V}_{t}^{\dagger} \nonumber\\
    &=\int \dd\mathcal{G}_{\operatorname{T}}\bigoplus_{k=1}^{p}v_t^k\rho^E\bigoplus_{k=1}^{p}v_t^{k\dagger}\nonumber\\
    &= \sum_{k=1}^{p} \int \dd\mu_k\mathcal{I}_{k}\left(v_{t}^{k}\right) \mathcal{I}_{k}(\varrho_{k}^{E}) \mathcal{I}_{k}\left(v_{t}^{k \dagger}\right) \nonumber\\
&= \sum_{k=1}^{p} \mathcal{I}_{k}\left( \int \dd\mu_k v_{t}^{k} \varrho_{k}^{E} v_{t}^{k \dagger}\right) \label{ls-1}\\
&= \bigoplus_{k=1}^{p} \frac{\operatorname{Tr}\{\varrho_{k}^{E}\}}{\dim{\{ \mathcal{H}_{k}\}} } \mathbb{1}_{n_t^k}. \label{ls-f}
\end{align}
\indent  Where Eq.~\eqref{ls-1} follows from the linearity of the integral and Eq.~\eqref{ls-f} follows from Eq.~\eqref{ef}. Furthermore, $\varrho_{k}^{E}$ is an $n_t^k \times n_t^k$ matrix corresponding to the part of $\rho^E$ in the subspace $\mathcal{H}_k$. Indeed, it is easy to see $\operatorname{Tr}\{\varrho_k^E\} = \operatorname{Tr}\{\rho_{n_t^k}^E\}$. Then
\begin{align}
     \Lambda_{\mathcal{G}_{\operatorname{T}}}[\rho^E]    &= \bigoplus_{k=1}^{p} \frac{\operatorname{Tr}\{\rho_{n_t^k}^E\}}{\dim{\{ \mathcal{H}_{k}\}} } \mathbb{1}_{n_t^k} \label{haar-final}
\end{align}
where $\rho_{n_t^k}\equiv \Pi_{n_t^k}\rho\Pi_{n_t^k}$ is the projection of $\rho$ into the subspace $\mathcal{H}_{k}$ which dimension is $\dim\{\mathcal{H}_k\} = n_t^k$. Therefore, this justifies the operator $\rho_{dd}^{E}$ as expressed in Eq.~\eqref{rhodd}. It is important to note that the elements of $\rho_{dd}^{E}$ are purely diagonal and are given by:

\begin{equation}
    c^{E}_{kk} = \frac{\operatorname{Tr}\{\rho_{n_t^k}^E\}}{\dim{\{ \mathcal{H}_{k}\}} } \label{elements-rhodd}
\end{equation}
which each $c_{kk}^{E}$ are associated to respectivity degenerate subspace $\mathcal{H}_{k}$.

Finally, since the Haar average depends only on the diagonal elements of the operator, the following equality is satisfied:
\begin{align}
     \Lambda_{\mathcal{G}_{\operatorname{T}}}[\rho^E]    &=      \Lambda_{\mathcal{G}_{\operatorname{T}}}[\rho^E_{diag}] \label{Haar-relation}
\end{align}
 where $\rho^E_{diag}$ is the diagonal part of the operator  $\rho$. Note that, Eq.~\eqref{Haar-relation} holds if and only if $\rho^E$ is not invariant under the thermodynamic group.

Therefore, using Eq.~\eqref{haar-final} we can compute the expression for the invariant work in Eq.~\eqref{inv-work-qpt}. In order to do this, let us consider the Hamiltonian in Eq.~\eqref{TdH}. Then
 \begin{align*}
W_{inv}\left[\rho\right] &=\int \dd\mathcal{G}_{\operatorname{T}}\int_{0}^{\tau+\epsilon} \dd t \operatorname{Tr}\left\{V_{t} \rho(t) V_{t}^{\dagger} \dfrac{d}{d t} H_g(t)\right\} \\
&=\dfrac{1}{2}\int \dd\mathcal{G}_{\operatorname{T}}\operatorname{Tr}\left\{V_{\tau} \rho(\tau) V_{\tau}^{\dagger} \left[H_g(\tau) -  H_{g_0}\right] \right\} \\
&=\dfrac{1}{2}\operatorname{Tr}\left\{\rho(\tau) H_g(\tau) - \Lambda_{\mathcal{G}_{\operatorname{T}}}[\rho^{E}(\tau)] H_{g_0}\right\} \\
&=\dfrac{1}{2}\operatorname{Tr}\left\{\rho^{E}_{diag}(\tau) H_g(\tau) - \rho_{dd}^{E}(\tau) H_{g_0}\right\}
 \end{align*}
where $\epsilon>0$ is arbitrarily small. Note that if we do not modify the Hamiltonian at $t=\tau$ by $\delta g H_1 = H(\tau)-H_{g_0}$, the above expression reduces to
\begin{equation}
W_{inv}[\rho] = \dfrac{\delta g}{2}\operatorname{Tr}\{\rho_{dd}^{E}(\tau)H_1\}. \label{new-winv}
\end{equation}
\indent The expression of invariant heat is obtained similarly. Furthermore, using Eq.~\eqref{new-winv} we can rewrite the coherent heat as
\begin{equation}
     Q_{c}[\rho] = \dfrac{\delta g}{2} \operatorname{Tr}\left\{\left(\rho^E(\tau) - \rho_{dd}^{E}(\tau)\right)H_1\right\}\label{new-coherent}
\end{equation}

In the particular case where the Hamiltonian exhibits no degeneracy, i. e. $n_k=1$ for all $k$, we obtain
\begin{equation*}
\rho_{dd}^{E}(\tau) =\displaystyle\bigoplus_{k=1}^{ d} \dfrac{\operatorname{Tr} \left\{\rho_{n_t^k}^{E}(\tau)\right \}}{1} \mathbb{1}_{1} = \rho^{E}_{diag}(\tau),
\end{equation*}
which is the density operator with the off-diagonal (in the the energy eigenbasis) elements removed. In this scenario, the invariant work reduces to
\begin{equation}
    W_{inv}[\rho] = \dfrac{1}{2} \operatorname{Tr}\left\{\left( H_g(\tau) - H_{g_0}\right)\rho_{diag}^{E}(\tau) \right\}, \label{particular-invariant-work}
\end{equation}
and the coherent heat takes the form
\begin{equation}
    Q_c[\rho] =  -\dfrac{1}{2}\operatorname{Tr}\left\{ \rho_c^{E}(\tau) H_{g_0}\right\}.
\end{equation}
\indent This demonstrates, in this case, that the term $Q_c[\rho]$ contains only contributions from the coherences. 

\section{\texorpdfstring{$\mathcal{G}_{\operatorname{T}}$-Entropy }{gt-entropy }}
\label{g-entropy}

Here we prove some results associated to the $\mathcal{G}_{\operatorname{T}}$-entropy. 

Using Eq.~\eqref{Haar-relation} and since $\Lambda_{\mathcal{G}_{\operatorname{T}}}$ is a unital quantum channel \cite{quantumt-twirling1,nielsen-channel,quantum-channel}, it follows from the data processing inequality \cite{cover,wilde} that 
\begin{equation}
    S_{\mathcal{G}}[\rho(t)] \geq S_{d}[\rho(t)] \label{first-inequality}
\end{equation}
for all $t$. This implies that there exists a non-negativity quantity such that $S_{\mathcal{G}}[\rho(t)] =  S_{d}[\rho(t)] + S_{\Gamma}[f_t]$. In this sense, from the Eq.~\eqref{Haar-relation} we obtain  the following inequality 
\begin{equation}
     S_{\Gamma}[f_t] =S_u[\Lambda_{\mathcal{G}_{\operatorname{T}}}(\rho^E_{diag})] -S_u[\rho^E_{diag}] \geq 0.\label{holevo-2}
\end{equation}
\indent \indent This shows that $S_{\Gamma}$ is, in fact, a Holevo asymmetry measure associated with the diagonal part of the density operator in the energy basis under uniform quantum twirling \cite{marvian}. Furthermore, as a consequence of the data processing inequality, the inequality in \eqref{holevo-2} is saturated if and only if $\Lambda_{\mathcal{G}_{\operatorname{T}}}(\rho^E) = \rho^E_{diag}$ \cite{wilde}, which occurs when $\rho^E$ is an invariant state or if the Hamiltonian has no degeneracies. Therefore, the diagonal entropy is equal to $S_{\mathcal{G}_{\operatorname{T}}}$ if and only if the Hamiltonian is not degenerate. Consequently, in this case, the state is symmetric with respect to the thermodynamic group and the Holevo asymmetry measure is zero, i.e. $ S_{\Gamma}[f_t] = 0$.

In the case where at least one energy level is non-degenerate   we can write the $\mathcal{G}_{\operatorname{T}}$-entropy as
\begin{align}
    -S_{\mathcal{G}_{\operatorname{T}}}[\rho] &= 
    \sum_{n=1}^d c_{nn}^{E}\log[c_{nn}^{E}] \nonumber\\
    &= \sum_m c_{mm}^{E}\log[c_{mm}^{E}] + \sum_{k} c_{kk}^{E}\log[c_{kk}^{E}], \label{sg-1}
\end{align}
 where $c_{kk}^{E}$ are given in Eq.~\eqref{elements-rhodd}.  The labels $m$ and $k$ refer to the average elements of $\rho^E$ associated with the $\mathcal{U}(1)$ group and the $\mathcal{U}(n_t^k>1)$ group, respectively. 

On the other hand,  we can split the diagonal entropy: $-S_d[\rho]  = \sum_{n=1}^{d} \rho^{E}_{n}\log[\rho^{E}_{n}] $ as
\begin{equation}
    -S_d[\rho] = \sum_m \rho^{E}_{mm}\log[\rho^{E}_{mm}]  +\sum_k \rho^E_{kk}\log[\rho^E_{kk}].\label{sd-entropy-split}
\end{equation}
\indent \indent Then, substituting Eq.~\eqref{sd-entropy-split} into Eq.~\eqref{sg-1}, we obtain the following expression:
\begin{align}
S_{\mathcal{G}_{\operatorname{T}}}[\rho]  &= S_d[\rho] \; + \nonumber \\
&+\sum_k \rho^E_{kk}\log[\rho^E_{kk}] - c_{kk}^{E}\log[c_{kk}^{E}] \label{sg-2}
\end{align}
\indent Now we introduce the functional $S_{\Gamma}[f_t] = -\operatorname{Tr}\{f_t\log(|f_t|)\}$  and define $f_t$ by 
\begin{equation}
    f_t\equiv  \left(\bigoplus_k -\rho^E_{kk} \mathbb{1}_{1} \right)\bigoplus \left(\bigoplus_{k}  \dfrac{\operatorname{Tr}\{\rho_{n_t^k}^E\}}{n_t^k} \mathbb{1}_{n_t^k}\right). \label{sg-3}
\end{equation}

\indent \indent In fact, $f_t$ is a diagonal matrix composed of two diagonal blocks. The first block is associated with the elements of the density operator in the energy basis, while the second block consists of the Haar average terms $c_{kk}^E$, in their explicit form given by Eq.~\eqref{elements-rhodd}. The elements in both blocks are associated with the degenerate subspaces where the Haar averages are related to the unitary group $\mathcal{U}(n_t^k>1)$. Consequently, using Eq.~\eqref{sg-2}, the functional $S_{\Gamma}[f_t]$ and $f_t$ given by Eq.~\eqref{sg-3}, Eq.~\eqref{connections-entropy} of the main text follow. 

Note that, if we consider the state $\sigma(t) \equiv D(\rho) = u_t\rho_{dd}^{E}u_t^{\dagger}$, we can show that the diagonal entropy of $\sigma(t)$ is
\begin{equation}
    S_{d}[\sigma(t)]  = S_{u}[\rho_{dd}^E] = S_{\mathcal{G}_{\operatorname{T}}}[\rho(t)].
\end{equation}

Therefore, for any invariant state $\rho_{dd}^E(t)$, the gauge entropy can be mapped to the diagonal entropy of a corresponding state $\sigma(t)$. This demonstrates that all thermodynamic properties of the diagonal entropy are extended to $S_{\mathcal{G}_{\operatorname{T}}}$.

\section{Applications}
\label{Appendix-B}
\subsection{Landau-Zener model}
\label{Appendix-B1}
Here we present some mathematical details concerning the invariant work in Eq.~\eqref{winv-explz} and the invariant heat in Eq.~\eqref{qc-explz} for the Landau-Zener model. 

First, we define, for simplicity, $\gamma_0 \equiv -\Delta/2 + ag_0$. For $t \geq \tau$, the Hamiltonian of the Landau-Zener model is given by Eq.~\eqref{Landau-Zener2}, which can be exactly diagonalized, resulting in the eigenvalues $E_{0,1}(g_0) = \pm \lambda =\pm \sqrt{(a \delta g+ \gamma_0)^2+\epsilon ^2}$ and respective normalized eigenvectors
\begin{equation}
\ket{a_{0}^{g}} = \begin{pmatrix}
-\dfrac{\epsilon}{\sqrt{\phi^2+\epsilon^2}} \\
\dfrac{\phi}{\sqrt{\phi^2+\epsilon^2}} \\
\end{pmatrix}, \;
\ket{a_{1}^{g}} = \begin{pmatrix}
\dfrac{\phi}{\sqrt{\phi^2+\epsilon^2}} \\
\dfrac{\epsilon}{\sqrt{\phi^2+\epsilon^2}} 
\end{pmatrix} 
\end{equation}
where $\phi \equiv \left( \lambda +a \delta g+\gamma_0\right)$.

In each process with $g_0\to g_0+\delta g$,  the initial Hamiltonian of the system is  $H_{g_0}=\gamma_0\sigma^z+\epsilon \sigma^x$ and we take the initial state of the system as $\rho\equiv \rho(0) = \ket{a_{0}^{g_0}}\bra{a_{0}^{g_0}}$, that is
\begin{align}
   \rho =
\begin{pmatrix}
 \dfrac{\sqrt{\epsilon ^2 +\gamma_0^2} - \gamma_0}{2 \sqrt{\gamma_0^2+\epsilon ^2}} & -\dfrac{\epsilon }{2 \sqrt{\gamma_0^2+\epsilon ^2}} \\
 -\dfrac{\epsilon }{2 \sqrt{\gamma_0^2+\epsilon ^2}} &  \dfrac{\sqrt{\epsilon ^2 +\gamma_0^2} + \gamma_0}{2\sqrt{\gamma_0^2+\epsilon ^2}}\\
\end{pmatrix}.
\end{align}
\indent \indent In the energy eigenbasis, we have $\rho^{E}$ transformed by the unitary matrix defined by $\ket{a_{0}^{g}}$ and $\ket{a_{1}^{g}}$. Its elements in this basis are given by
\begin{align}
    \begin{cases}
        \rho^E_{11} =   \dfrac{\lambda\sqrt{\gamma_0^2+\epsilon ^2} + ( (\phi - \lambda)\gamma_0 +\epsilon ^2)}{2 \lambda \sqrt{\gamma_0^2+\epsilon ^2} }   \\
        \rho^E_{12} =   \rho^E_{21} =-\dfrac{a \delta g \epsilon }{2 \lambda\sqrt{\gamma_0^2+\epsilon ^2}} \\ 
        \rho^E_{22} =\dfrac{\lambda\sqrt{\gamma_0^2+\epsilon ^2} -( (\phi - \lambda)\gamma_0 +\epsilon ^2)}{2 \lambda \sqrt{\gamma_0^2+\epsilon ^2} } 
    \end{cases}. \label{densitys}
\end{align}
\indent \indent Therefore, using the expressions in Eq.~\eqref{densitys} we obtain the diagonal part of $\rho$, in energ eigenbasis, i.e $\rho_{diag}^E = \operatorname{diag}(\rho^E_{11} \; \; \rho^E_{22} )$. For the coherent density matrix, in energy eigenbasis, we can compute $\rho_c^E = \rho^{E}  - \rho_{diag}^E$, then these elements are  defined  by $(\rho_{c}^E)_{jk} = (\rho_{jk}^E)_{jk}$ if $j\neq k$ and $(\rho_{c}^E)_{jk} = 0$ if $j=k$.

Now, we need to transform each Hamiltonian to the energy basis for $t \geq \tau$. IIn this basis,  $H(\tau)$ simply the diagonal matrix with its eigenvalues  i.e $H^{E}(\tau) = \operatorname{diag}(-\lambda \; \lambda)$. Then  we only need to obtain the representation of $H_0$, which is given by
\begin{align}
   H^E_{g_{0}} =  \begin{pmatrix}
   \dfrac{(\lambda -\phi)\gamma_0- \epsilon^2}{\lambda} &  \dfrac{a\epsilon \delta g}{\lambda} \\
        \dfrac{a\epsilon \delta g}{\lambda} &  \dfrac{(\phi - \lambda)\gamma_0 + \epsilon^2}{\lambda}
   \end{pmatrix} \label{trh0}
\end{align}
\indent We can now compute the invariant work and heat. Making it term by term, we have
\begin{align}
\begin{cases}
     \operatorname{Tr}\left\{\rho_{diag}^{E} H_{g_{0}}^{E} \right\} =   -\dfrac{\left( (\phi - \lambda)\gamma_0 +\epsilon ^2\right)^2}{\lambda\sqrt{\gamma_0^2+\epsilon ^2} } \\
    \operatorname{Tr}\left\{\rho_{diag}^{E} H^{E}(\tau) \right\} =  -\dfrac{(\phi - \lambda)\gamma_0 +\epsilon ^2}{\sqrt{\gamma_0^2+\epsilon ^2}} \\
    \operatorname{Tr}\{\rho_c^EH_0^E\} = -\dfrac{a^2 g^2 \epsilon ^2}{\lambda \sqrt{\gamma_0^2+\epsilon ^2} }. 
\end{cases} \label{tr1}
\end{align}

\indent Using the  expressions in Eq.~\eqref{tr1}  we can immediately obtain the we immediately obtain the expressions for the invariant work and coherent heat given in Eqs.~\eqref{winv-explz} and~\eqref{qc-explz} of the main text.

Then, from Eqs.~\eqref{winv-explz} and~\eqref{qc-explz}, we obtain their derivatives with respect to $g_0$ given by
\begin{widetext}
    \begin{align}    
 \frac{dW_{inv}[\rho]}{dg_0}&=    -\frac{a^2 \delta g \epsilon ^2 \left(a^4 \delta g^4+3 a^3 \delta g^3 \gamma_0+2 a^2 \delta g^2 \gamma_0^2 + a \delta g \gamma_0 \left(\gamma_0^2+\epsilon ^2\right)+\left(\gamma_0^2+\epsilon ^2\right)^2\right)}{2 \left(\gamma_0^2+\epsilon ^2\right)^{3/2} \left((a \delta g+\gamma_0)^2+\epsilon ^2\right)^2} \\  
\frac{dQ_c[\rho]}{dg_0}&= -\frac{a^2 \delta g^2 \epsilon ^2 \left(\epsilon ^2 (2 a \delta g+3 \gamma_0)+\gamma_0 (a \delta g+\gamma_0) (a \delta g+3 \gamma_0)\right)}{2 \left(\gamma_0^2+\epsilon ^2\right)^{3/2} \left((a \delta g+\gamma_0)^2+\epsilon ^2\right)^2}
\end{align}
\end{widetext}
\indent \indent Furthermore, diagonal entropy can be computed using $\rho^E_{11}$ and $\rho^E_{22}$.

Finally, let us consider the invariant work and its connections to Hellman-Feynman Theorem~\cite{Feynman}. In fact, from this theorem it follows
\begin{equation}
    \dfrac{dE_0(g)}{dg} = \operatorname{Tr}\left\{\rho H_1 \right\} = \dfrac{2}{g} W_{inv}[\rho] + \dfrac{2}{g} Q_c[\rho]\label{eq-hellman-feynman}
\end{equation} 
since $g\neq 0$, $H_1 = (1/g)(H(\tau)  - H_0)$ and $\rho = \rho_{diag} + \rho_c$. This result is valid if and only if the spectrum of the Hamiltonian is not degenerated. In the particular case where $Q_c[\rho]\approx 0$, Eq.~\eqref{eq-hellman-feynman} have only contributions from the invariant work.


\providecommand{\noopsort}[1]{}\providecommand{\singleletter}[1]{#1}%
\begin{thebibliography}{66}%
\makeatletter
\providecommand \@ifxundefined [1]{%
 \@ifx{#1\undefined}
}%
\providecommand \@ifnum [1]{%
 \ifnum #1\expandafter \@firstoftwo
 \else \expandafter \@secondoftwo
 \fi
}%
\providecommand \@ifx [1]{%
 \ifx #1\expandafter \@firstoftwo
 \else \expandafter \@secondoftwo
 \fi
}%
\providecommand \natexlab [1]{#1}%
\providecommand \enquote  [1]{``#1''}%
\providecommand \bibnamefont  [1]{#1}%
\providecommand \bibfnamefont [1]{#1}%
\providecommand \citenamefont [1]{#1}%
\providecommand \href@noop [0]{\@secondoftwo}%
\providecommand \href [0]{\begingroup \@sanitize@url \@href}%
\providecommand \@href[1]{\@@startlink{#1}\@@href}%
\providecommand \@@href[1]{\endgroup#1\@@endlink}%
\providecommand \@sanitize@url [0]{\catcode `\\12\catcode `\$12\catcode `\&12\catcode `\#12\catcode `\^12\catcode `\_12\catcode `\%12\relax}%
\providecommand \@@startlink[1]{}%
\providecommand \@@endlink[0]{}%
\providecommand \url  [0]{\begingroup\@sanitize@url \@url }%
\providecommand \@url [1]{\endgroup\@href {#1}{\urlprefix }}%
\providecommand \urlprefix  [0]{URL }%
\providecommand \Eprint [0]{\href }%
\providecommand \doibase [0]{https://doi.org/}%
\providecommand \selectlanguage [0]{\@gobble}%
\providecommand \bibinfo  [0]{\@secondoftwo}%
\providecommand \bibfield  [0]{\@secondoftwo}%
\providecommand \translation [1]{[#1]}%
\providecommand \BibitemOpen [0]{}%
\providecommand \bibitemStop [0]{}%
\providecommand \bibitemNoStop [0]{.\EOS\space}%
\providecommand \EOS [0]{\spacefactor3000\relax}%
\providecommand \BibitemShut  [1]{\csname bibitem#1\endcsname}%
\let\auto@bib@innerbib\@empty
\bibitem [{\citenamefont {Goold}\ \emph {et~al.}(2016)\citenamefont {Goold}, \citenamefont {Huber}, \citenamefont {Riera}, \citenamefont {del Rio},\ and\ \citenamefont {Skrzypczyk}}]{Goold2016}%
  \BibitemOpen
  \bibfield  {author} {\bibinfo {author} {\bibfnamefont {J.}~\bibnamefont {Goold}}, \bibinfo {author} {\bibfnamefont {M.}~\bibnamefont {Huber}}, \bibinfo {author} {\bibfnamefont {A.}~\bibnamefont {Riera}}, \bibinfo {author} {\bibfnamefont {L.}~\bibnamefont {del Rio}},\ and\ \bibinfo {author} {\bibfnamefont {P.}~\bibnamefont {Skrzypczyk}},\ }\bibfield  {title} {\bibinfo {title} {The role of quantum information in thermodynamicsa topical review},\ }\href@noop {} {\bibfield  {journal} {\bibinfo  {journal} {Journal of Physics A: Mathematical and Theoretical}\ }\textbf {\bibinfo {volume} {49}},\ \bibinfo {pages} {143001} (\bibinfo {year} {2016})}\BibitemShut {NoStop}%
\bibitem [{\citenamefont {Callen}(1985)}]{Callen1985}%
  \BibitemOpen
  \bibfield  {author} {\bibinfo {author} {\bibfnamefont {H.~B.}\ \bibnamefont {Callen}},\ }\href@noop {} {\emph {\bibinfo {title} {Thermodynamics and an Introduction to Thermostatistics}}}\ (\bibinfo  {publisher} {Wiley},\ \bibinfo {address} {Hoboken, NJ, USA},\ \bibinfo {year} {1985})\BibitemShut {NoStop}%
\bibitem [{\citenamefont {Di~Bella}(2021)}]{Bella2021}%
  \BibitemOpen
  \bibfield  {author} {\bibinfo {author} {\bibfnamefont {F.}~\bibnamefont {Di~Bella}},\ }\href@noop {} {\emph {\bibinfo {title} {Applying Engineering Thermodynamics}}}\ (\bibinfo  {publisher} {World Scientific Publishing Company},\ \bibinfo {address} {Hong Kong},\ \bibinfo {year} {2021})\BibitemShut {NoStop}%
\bibitem [{\citenamefont {Mizraji}(2024)}]{Mizraji2024}%
  \BibitemOpen
  \bibfield  {author} {\bibinfo {author} {\bibfnamefont {E.}~\bibnamefont {Mizraji}},\ }\bibfield  {title} {\bibinfo {title} {Homeostasis and information processing: The key frames for the thermodynamics of biological systems},\ }\href {https://doi.org/https://doi.org/10.1016/j.biosystems.2023.105115} {\bibfield  {journal} {\bibinfo  {journal} {Biosystems}\ }\textbf {\bibinfo {volume} {236}},\ \bibinfo {pages} {105115} (\bibinfo {year} {2024})}\BibitemShut {NoStop}%
\bibitem [{\citenamefont {Deli}\ \emph {et~al.}(2021)\citenamefont {Deli}, \citenamefont {Peters},\ and\ \citenamefont {Kisv\'{a}rday}}]{Deli2021}%
  \BibitemOpen
  \bibfield  {author} {\bibinfo {author} {\bibfnamefont {E.}~\bibnamefont {Deli}}, \bibinfo {author} {\bibfnamefont {J.}~\bibnamefont {Peters}},\ and\ \bibinfo {author} {\bibfnamefont {Z.}~\bibnamefont {Kisv\'{a}rday}},\ }\bibfield  {title} {\bibinfo {title} {The thermodynamics of cognition: A mathematical treatment},\ }\href {https://doi.org/https://doi.org/10.1016/j.csbj.2021.01.008} {\bibfield  {journal} {\bibinfo  {journal} {Computational and Structural Biotechnology Journal}\ }\textbf {\bibinfo {volume} {19}},\ \bibinfo {pages} {784} (\bibinfo {year} {2021})}\BibitemShut {NoStop}%
\bibitem [{\citenamefont {Gross}(2001)}]{Gross2001}%
  \BibitemOpen
  \bibfield  {author} {\bibinfo {author} {\bibfnamefont {D.~H.~E.}\ \bibnamefont {Gross}},\ }\href@noop {} {\emph {\bibinfo {title} {Microcanonical thermodynamics: phase transitions in "small" systems}}},\ Vol.~\bibinfo {volume} {66}\ (\bibinfo  {publisher} {World Scientific},\ \bibinfo {year} {2001})\BibitemShut {NoStop}%
\bibitem [{\citenamefont {Bennett}(1982)}]{Bennett1982}%
  \BibitemOpen
  \bibfield  {author} {\bibinfo {author} {\bibfnamefont {C.~H.}\ \bibnamefont {Bennett}},\ }\bibfield  {title} {\bibinfo {title} {The thermodynamics of computation—a review},\ }\href@noop {} {\bibfield  {journal} {\bibinfo  {journal} {International Journal of Theoretical Physics}\ }\textbf {\bibinfo {volume} {21}},\ \bibinfo {pages} {905} (\bibinfo {year} {1982})}\BibitemShut {NoStop}%
\bibitem [{\citenamefont {Wald}(2001)}]{Wald2001}%
  \BibitemOpen
  \bibfield  {author} {\bibinfo {author} {\bibfnamefont {R.~M.}\ \bibnamefont {Wald}},\ }\bibfield  {title} {\bibinfo {title} {The thermodynamics of black holes},\ }\href@noop {} {\bibfield  {journal} {\bibinfo  {journal} {Living Rev. Relativ.}\ }\textbf {\bibinfo {volume} {4}},\ \bibinfo {pages} {1} (\bibinfo {year} {2001})}\BibitemShut {NoStop}%
\bibitem [{\citenamefont {Strasberg}(2022)}]{Strasberg2022}%
  \BibitemOpen
  \bibfield  {author} {\bibinfo {author} {\bibfnamefont {P.}~\bibnamefont {Strasberg}},\ }\href@noop {} {\emph {\bibinfo {title} {Quantum Stochastic Thermodynamics: Foundations and Selected Applications}}}\ (\bibinfo  {publisher} {Oxford University Press},\ \bibinfo {address} {Oxford, UK},\ \bibinfo {year} {2022})\BibitemShut {NoStop}%
\bibitem [{\citenamefont {Lieb}\ and\ \citenamefont {Yngvason}(1999)}]{Lieb1999}%
  \BibitemOpen
  \bibfield  {author} {\bibinfo {author} {\bibfnamefont {E.~H.}\ \bibnamefont {Lieb}}\ and\ \bibinfo {author} {\bibfnamefont {J.}~\bibnamefont {Yngvason}},\ }\bibfield  {title} {\bibinfo {title} {The physics and mathematics of the second law of thermodynamics},\ }\href@noop {} {\bibfield  {journal} {\bibinfo  {journal} {Physics Reports}\ }\textbf {\bibinfo {volume} {310}},\ \bibinfo {pages} {1} (\bibinfo {year} {1999})}\BibitemShut {NoStop}%
\bibitem [{\citenamefont {Giles}(2016)}]{Giles2016}%
  \BibitemOpen
  \bibfield  {author} {\bibinfo {author} {\bibfnamefont {R.}~\bibnamefont {Giles}},\ }\href@noop {} {\emph {\bibinfo {title} {Mathematical foundations of thermodynamics: International series of monographs on pure and applied mathematics}}},\ Vol.~\bibinfo {volume} {53}\ (\bibinfo  {publisher} {Elsevier},\ \bibinfo {year} {2016})\BibitemShut {NoStop}%
\bibitem [{\citenamefont {Alicki}(1979)}]{Alicki1979}%
  \BibitemOpen
  \bibfield  {author} {\bibinfo {author} {\bibfnamefont {R.}~\bibnamefont {Alicki}},\ }\bibfield  {title} {\bibinfo {title} {The quantum open system as a model of the heat engine},\ }\href@noop {} {\bibfield  {journal} {\bibinfo  {journal} {Journal of Physics A: Mathematical and General}\ }\textbf {\bibinfo {volume} {12}},\ \bibinfo {pages} {L103} (\bibinfo {year} {1979})}\BibitemShut {NoStop}%
\bibitem [{\citenamefont {Alicki}\ and\ \citenamefont {Kosloff}(2018)}]{Alicki2018}%
  \BibitemOpen
  \bibfield  {author} {\bibinfo {author} {\bibfnamefont {R.}~\bibnamefont {Alicki}}\ and\ \bibinfo {author} {\bibfnamefont {R.}~\bibnamefont {Kosloff}},\ }\bibinfo {title} {Introduction to quantum thermodynamics: History and prospects},\ in\ \href {https://doi.org/10.1007/978-3-319-99046-0_1} {\emph {\bibinfo {booktitle} {Thermodynamics in the Quantum Regime: Fundamental Aspects and New Directions}}}\ (\bibinfo  {publisher} {Springer International Publishing},\ \bibinfo {address} {Cham},\ \bibinfo {year} {2018})\ pp.\ \bibinfo {pages} {1--33}\BibitemShut {NoStop}%
\bibitem [{\citenamefont {Binder}\ \emph {et~al.}(2018)\citenamefont {Binder}, \citenamefont {Correa}, \citenamefont {Gogolin}, \citenamefont {Anders},\ and\ \citenamefont {Adesso}}]{Binder2018}%
  \BibitemOpen
  \bibinfo {editor} {\bibfnamefont {F.}~\bibnamefont {Binder}}, \bibinfo {editor} {\bibfnamefont {L.~A.}\ \bibnamefont {Correa}}, \bibinfo {editor} {\bibfnamefont {C.}~\bibnamefont {Gogolin}}, \bibinfo {editor} {\bibfnamefont {J.}~\bibnamefont {Anders}},\ and\ \bibinfo {editor} {\bibfnamefont {G.}~\bibnamefont {Adesso}},\ eds.,\ \href {https://doi.org/10.1007/978-3-319-99046-0} {\emph {\bibinfo {title} {Thermodynamics in the Quantum Regime: Fundamental Aspects and New Directions}}}\ (\bibinfo  {publisher} {Springer International Publishing},\ \bibinfo {address} {Cham},\ \bibinfo {year} {2018})\BibitemShut {NoStop}%
\bibitem [{\citenamefont {Kurchan}(2000)}]{Kurchan2000}%
  \BibitemOpen
  \bibfield  {author} {\bibinfo {author} {\bibfnamefont {J.}~\bibnamefont {Kurchan}},\ }\bibfield  {title} {\bibinfo {title} {A quantum fluctuation theorem},\ }\href@noop {} {\bibfield  {journal} {\bibinfo  {journal} {arXiv preprint arXiv:0007360}\ } (\bibinfo {year} {2000})}\BibitemShut {NoStop}%
\bibitem [{\citenamefont {Talkner}\ \emph {et~al.}(2007)\citenamefont {Talkner}, \citenamefont {Lutz},\ and\ \citenamefont {H{\"a}nggi}}]{Talkner2007}%
  \BibitemOpen
  \bibfield  {author} {\bibinfo {author} {\bibfnamefont {P.}~\bibnamefont {Talkner}}, \bibinfo {author} {\bibfnamefont {E.}~\bibnamefont {Lutz}},\ and\ \bibinfo {author} {\bibfnamefont {P.}~\bibnamefont {H{\"a}nggi}},\ }\bibfield  {title} {\bibinfo {title} {Fluctuation theorems: Work is not an observable},\ }\href@noop {} {\bibfield  {journal} {\bibinfo  {journal} {Physical Review E}\ }\textbf {\bibinfo {volume} {75}},\ \bibinfo {pages} {050102} (\bibinfo {year} {2007})}\BibitemShut {NoStop}%
\bibitem [{\citenamefont {Horowitz}\ and\ \citenamefont {Parrondo}(2012)}]{Horowitz2012}%
  \BibitemOpen
  \bibfield  {author} {\bibinfo {author} {\bibfnamefont {J.~M.}\ \bibnamefont {Horowitz}}\ and\ \bibinfo {author} {\bibfnamefont {J.~M.~R.}\ \bibnamefont {Parrondo}},\ }\bibfield  {title} {\bibinfo {title} {Entropy production along nonequilibrium quantum jump trajectories},\ }\href@noop {} {\bibfield  {journal} {\bibinfo  {journal} {New Journal of Physics}\ }\textbf {\bibinfo {volume} {14}},\ \bibinfo {pages} {123019} (\bibinfo {year} {2012})}\BibitemShut {NoStop}%
\bibitem [{\citenamefont {Lostaglio}\ \emph {et~al.}(2015)\citenamefont {Lostaglio}, \citenamefont {Jennings},\ and\ \citenamefont {Rudolph}}]{Lostaglio2015}%
  \BibitemOpen
  \bibfield  {author} {\bibinfo {author} {\bibfnamefont {M.}~\bibnamefont {Lostaglio}}, \bibinfo {author} {\bibfnamefont {D.}~\bibnamefont {Jennings}},\ and\ \bibinfo {author} {\bibfnamefont {T.}~\bibnamefont {Rudolph}},\ }\bibfield  {title} {\bibinfo {title} {Description of quantum coherence in thermodynamic processes requires constraints beyond free energy},\ }\href@noop {} {\bibfield  {journal} {\bibinfo  {journal} {Nature Communications}\ }\textbf {\bibinfo {volume} {6}},\ \bibinfo {pages} {6383} (\bibinfo {year} {2015})}\BibitemShut {NoStop}%
\bibitem [{\citenamefont {Talkner}\ and\ \citenamefont {H{\"a}nggi}(2016)}]{Talkner2016}%
  \BibitemOpen
  \bibfield  {author} {\bibinfo {author} {\bibfnamefont {P.}~\bibnamefont {Talkner}}\ and\ \bibinfo {author} {\bibfnamefont {P.}~\bibnamefont {H{\"a}nggi}},\ }\bibfield  {title} {\bibinfo {title} {Aspects of quantum work},\ }\href@noop {} {\bibfield  {journal} {\bibinfo  {journal} {Phys. Rev. E}\ }\textbf {\bibinfo {volume} {93}},\ \bibinfo {pages} {022131} (\bibinfo {year} {2016})}\BibitemShut {NoStop}%
\bibitem [{\citenamefont {von Neumann}(1927)}]{Neumann1927}%
  \BibitemOpen
  \bibfield  {author} {\bibinfo {author} {\bibfnamefont {J.}~\bibnamefont {von Neumann}},\ }\bibfield  {title} {\bibinfo {title} {Thermodynamik quantenmechanischer gesamtheiten},\ }\href@noop {} {\bibfield  {journal} {\bibinfo  {journal} {Nachrichten von der Gesellschaft der Wissenschaften zu Göttingen, Mathematisch-Physikalische Klasse}\ }\textbf {\bibinfo {volume} {1927}},\ \bibinfo {pages} {273} (\bibinfo {year} {1927})}\BibitemShut {NoStop}%
\bibitem [{\citenamefont {Polkovnikov}(2011)}]{Polkovnikov2011}%
  \BibitemOpen
  \bibfield  {author} {\bibinfo {author} {\bibfnamefont {A.}~\bibnamefont {Polkovnikov}},\ }\bibfield  {title} {\bibinfo {title} {Microscopic diagonal entropy and its connection to basic thermodynamic relations},\ }\href@noop {} {\bibfield  {journal} {\bibinfo  {journal} {Annals of Physics}\ }\textbf {\bibinfo {volume} {326}},\ \bibinfo {pages} {486} (\bibinfo {year} {2011})}\BibitemShut {NoStop}%
\bibitem [{\citenamefont {Kawai}\ \emph {et~al.}(2007)\citenamefont {Kawai}, \citenamefont {Parrondo},\ and\ \citenamefont {den Broeck}}]{Kawai2007}%
  \BibitemOpen
  \bibfield  {author} {\bibinfo {author} {\bibfnamefont {R.}~\bibnamefont {Kawai}}, \bibinfo {author} {\bibfnamefont {J.~M.}\ \bibnamefont {Parrondo}},\ and\ \bibinfo {author} {\bibfnamefont {C.~V.}\ \bibnamefont {den Broeck}},\ }\bibfield  {title} {\bibinfo {title} {Dissipation: The phase-space perspective},\ }\href@noop {} {\bibfield  {journal} {\bibinfo  {journal} {Physical review letters}\ }\textbf {\bibinfo {volume} {98}},\ \bibinfo {pages} {080602} (\bibinfo {year} {2007})}\BibitemShut {NoStop}%
\bibitem [{\citenamefont {Sagawa}(2013)}]{Sagawa2013}%
  \BibitemOpen
  \bibfield  {author} {\bibinfo {author} {\bibfnamefont {T.}~\bibnamefont {Sagawa}},\ }\bibfield  {title} {\bibinfo {title} {Second law-like inequalities with quantum relative entropy: An introduction},\ }in\ \href@noop {} {\emph {\bibinfo {booktitle} {Lectures on quantum computing, thermodynamics and statistical physics}}}\ (\bibinfo  {publisher} {World Scientific},\ \bibinfo {year} {2013})\ pp.\ \bibinfo {pages} {125--190}\BibitemShut {NoStop}%
\bibitem [{\citenamefont {Vedral}(2002)}]{Vedral2002}%
  \BibitemOpen
  \bibfield  {author} {\bibinfo {author} {\bibfnamefont {V.}~\bibnamefont {Vedral}},\ }\bibfield  {title} {\bibinfo {title} {The role of relative entropy in quantum information theory},\ }\href@noop {} {\bibfield  {journal} {\bibinfo  {journal} {Reviews of Modern Physics}\ }\textbf {\bibinfo {volume} {74}},\ \bibinfo {pages} {197} (\bibinfo {year} {2002})}\BibitemShut {NoStop}%
\bibitem [{\citenamefont {Landi}\ and\ \citenamefont {Paternostro}(2021)}]{Landi2021}%
  \BibitemOpen
  \bibfield  {author} {\bibinfo {author} {\bibfnamefont {G.~T.}\ \bibnamefont {Landi}}\ and\ \bibinfo {author} {\bibfnamefont {M.}~\bibnamefont {Paternostro}},\ }\bibfield  {title} {\bibinfo {title} {Irreversible entropy production: From classical to quantum},\ }\href@noop {} {\bibfield  {journal} {\bibinfo  {journal} {Reviews of Modern Physics}\ }\textbf {\bibinfo {volume} {93}},\ \bibinfo {pages} {035008} (\bibinfo {year} {2021})}\BibitemShut {NoStop}%
\bibitem [{\citenamefont {Weinberg}(1995)}]{Weinberg1995}%
  \BibitemOpen
  \bibfield  {author} {\bibinfo {author} {\bibfnamefont {S.}~\bibnamefont {Weinberg}},\ }\href@noop {} {\emph {\bibinfo {title} {The Quantum Theory of Fields, Vol. 1: Foundations}}}\ (\bibinfo  {publisher} {Cambridge University Press},\ \bibinfo {address} {Cambridge},\ \bibinfo {year} {1995})\BibitemShut {NoStop}%
\bibitem [{\citenamefont {Rovelli}(2004)}]{Rovelli2004}%
  \BibitemOpen
  \bibfield  {author} {\bibinfo {author} {\bibfnamefont {C.}~\bibnamefont {Rovelli}},\ }\href@noop {} {\emph {\bibinfo {title} {Quantum Gravity}}}\ (\bibinfo  {publisher} {Cambridge University Press},\ \bibinfo {address} {Cambridge},\ \bibinfo {year} {2004})\BibitemShut {NoStop}%
\bibitem [{\citenamefont {Céleri}\ and\ \citenamefont {Rudnicki}(2024)}]{gauge-lucas}%
  \BibitemOpen
  \bibfield  {author} {\bibinfo {author} {\bibfnamefont {L.~C.}\ \bibnamefont {Céleri}}\ and\ \bibinfo {author} {\bibfnamefont {L.}~\bibnamefont {Rudnicki}},\ }\bibfield  {title} {\bibinfo {title} {Gauge-invariant quantum thermodynamics: Consequences for the first law},\ }\href {https://doi.org/10.3390/e26020111} {\bibfield  {journal} {\bibinfo  {journal} {Entropy}\ }\textbf {\bibinfo {volume} {26}},\ \bibinfo {pages} {111} (\bibinfo {year} {2024})}\BibitemShut {NoStop}%
\bibitem [{\citenamefont {Ara{\'u}jo}\ \emph {et~al.}(2018)\citenamefont {Ara{\'u}jo}, \citenamefont {H{\"a}ffner}, \citenamefont {Bernardi}, \citenamefont {Tasca}, \citenamefont {Lavery}, \citenamefont {Padgett}, \citenamefont {Kanaan}, \citenamefont {C{\'e}leri},\ and\ \citenamefont {Souto~Ribeiro}}]{Araujo2018}%
  \BibitemOpen
  \bibfield  {author} {\bibinfo {author} {\bibfnamefont {R.~M.}\ \bibnamefont {Ara{\'u}jo}}, \bibinfo {author} {\bibfnamefont {T.}~\bibnamefont {H{\"a}ffner}}, \bibinfo {author} {\bibfnamefont {R.}~\bibnamefont {Bernardi}}, \bibinfo {author} {\bibfnamefont {D.~S.}\ \bibnamefont {Tasca}}, \bibinfo {author} {\bibfnamefont {M.~P.~J.}\ \bibnamefont {Lavery}}, \bibinfo {author} {\bibfnamefont {M.~J.}\ \bibnamefont {Padgett}}, \bibinfo {author} {\bibfnamefont {A.}~\bibnamefont {Kanaan}}, \bibinfo {author} {\bibfnamefont {L.~C.}\ \bibnamefont {C{\'e}leri}},\ and\ \bibinfo {author} {\bibfnamefont {P.~H.}\ \bibnamefont {Souto~Ribeiro}},\ }\bibfield  {title} {\bibinfo {title} {Experimental study of quantum thermodynamics using optical vortices},\ }\href@noop {} {\bibfield  {journal} {\bibinfo  {journal} {Journal of Physics Communications}\ }\textbf {\bibinfo {volume} {2}},\ \bibinfo {pages} {035012} (\bibinfo {year} {2018})}\BibitemShut {NoStop}%
\bibitem [{\citenamefont {Brandao}\ \emph {et~al.}(2013)\citenamefont {Brandao}, \citenamefont {Horodecki}, \citenamefont {Oppenheim}, \citenamefont {Renes},\ and\ \citenamefont {Spekkens}}]{Brandao2013}%
  \BibitemOpen
  \bibfield  {author} {\bibinfo {author} {\bibfnamefont {F.~G. S.~L.}\ \bibnamefont {Brandao}}, \bibinfo {author} {\bibfnamefont {M.}~\bibnamefont {Horodecki}}, \bibinfo {author} {\bibfnamefont {J.}~\bibnamefont {Oppenheim}}, \bibinfo {author} {\bibfnamefont {J.~M.}\ \bibnamefont {Renes}},\ and\ \bibinfo {author} {\bibfnamefont {R.~W.}\ \bibnamefont {Spekkens}},\ }\bibfield  {title} {\bibinfo {title} {Resource theory of quantum states out of thermal equilibrium},\ }\href@noop {} {\bibfield  {journal} {\bibinfo  {journal} {Physical Review Letters}\ }\textbf {\bibinfo {volume} {111}},\ \bibinfo {pages} {250404} (\bibinfo {year} {2013})}\BibitemShut {NoStop}%
\bibitem [{\citenamefont {Chitambar}\ and\ \citenamefont {Gour}(2019)}]{Chitambar2019}%
  \BibitemOpen
  \bibfield  {author} {\bibinfo {author} {\bibfnamefont {E.}~\bibnamefont {Chitambar}}\ and\ \bibinfo {author} {\bibfnamefont {G.}~\bibnamefont {Gour}},\ }\bibfield  {title} {\bibinfo {title} {Quantum resources theories},\ }\href@noop {} {\bibfield  {journal} {\bibinfo  {journal} {Reviews of Modern Physics}\ }\textbf {\bibinfo {volume} {91}},\ \bibinfo {pages} {025001} (\bibinfo {year} {2019})}\BibitemShut {NoStop}%
\bibitem [{\citenamefont {Fulton}\ and\ \citenamefont {Harris}(1991)}]{representation}%
  \BibitemOpen
  \bibfield  {author} {\bibinfo {author} {\bibfnamefont {W.}~\bibnamefont {Fulton}}\ and\ \bibinfo {author} {\bibfnamefont {J.}~\bibnamefont {Harris}},\ }\href {https://books.google.com.br/books?id=qGFzi20nMcYC} {\emph {\bibinfo {title} {Representation Theory: A First Course}}},\ Graduate Texts in Mathematics\ (\bibinfo  {publisher} {Springer New York},\ \bibinfo {year} {1991})\BibitemShut {NoStop}%
\bibitem [{\citenamefont {Mele}(2024)}]{haar-information}%
  \BibitemOpen
  \bibfield  {author} {\bibinfo {author} {\bibfnamefont {A.~A.}\ \bibnamefont {Mele}},\ }\bibfield  {title} {\bibinfo {title} {Introduction to haar measure tools in quantum information: A beginner's tutorial},\ }\href {https://doi.org/10.22331/q-2024-05-08-1340} {\bibfield  {journal} {\bibinfo  {journal} {Quantum}\ }\textbf {\bibinfo {volume} {8}},\ \bibinfo {pages} {1340} (\bibinfo {year} {2024})}\BibitemShut {NoStop}%
\bibitem [{\citenamefont {Bartlett}\ and\ \citenamefont {Wiseman}(2003)}]{quantumt-twirling1}%
  \BibitemOpen
  \bibfield  {author} {\bibinfo {author} {\bibfnamefont {S.~D.}\ \bibnamefont {Bartlett}}\ and\ \bibinfo {author} {\bibfnamefont {H.~M.}\ \bibnamefont {Wiseman}},\ }\bibfield  {title} {\bibinfo {title} {Entanglement constrained by superselection rules},\ }\bibfield  {journal} {\bibinfo  {journal} {Physical Review Letters}\ }\textbf {\bibinfo {volume} {91}},\ \href {https://doi.org/10.1103/physrevlett.91.097903} {10.1103/physrevlett.91.097903} (\bibinfo {year} {2003})\BibitemShut {NoStop}%
\bibitem [{\citenamefont {Dankert}\ \emph {et~al.}(2009)\citenamefont {Dankert}, \citenamefont {Cleve}, \citenamefont {Emerson},\ and\ \citenamefont {Livine}}]{quantum-channel}%
  \BibitemOpen
  \bibfield  {author} {\bibinfo {author} {\bibfnamefont {C.}~\bibnamefont {Dankert}}, \bibinfo {author} {\bibfnamefont {R.}~\bibnamefont {Cleve}}, \bibinfo {author} {\bibfnamefont {J.}~\bibnamefont {Emerson}},\ and\ \bibinfo {author} {\bibfnamefont {E.}~\bibnamefont {Livine}},\ }\bibfield  {title} {\bibinfo {title} {Exact and approximate unitary 2-designs and their application to fidelity estimation},\ }\href {https://doi.org/10.1103/PhysRevA.80.012304} {\bibfield  {journal} {\bibinfo  {journal} {Phys. Rev. A}\ }\textbf {\bibinfo {volume} {80}},\ \bibinfo {pages} {012304} (\bibinfo {year} {2009})}\BibitemShut {NoStop}%
\bibitem [{\citenamefont {Nielsen}\ and\ \citenamefont {Chuang}(2010)}]{nielsen}%
  \BibitemOpen
  \bibfield  {author} {\bibinfo {author} {\bibfnamefont {M.}~\bibnamefont {Nielsen}}\ and\ \bibinfo {author} {\bibfnamefont {I.}~\bibnamefont {Chuang}},\ }\href {https://books.google.com.br/books?id=-s4DEy7o-a0C} {\emph {\bibinfo {title} {Quantum Computation and Quantum Information: 10th Anniversary Edition}}}\ (\bibinfo  {publisher} {Cambridge University Press},\ \bibinfo {year} {2010})\BibitemShut {NoStop}%
\bibitem [{\citenamefont {Chinni}\ \emph {et~al.}(2021)\citenamefont {Chinni}, \citenamefont {Poggi},\ and\ \citenamefont {Deutsch}}]{LMG-QPT1}%
  \BibitemOpen
  \bibfield  {author} {\bibinfo {author} {\bibfnamefont {K.}~\bibnamefont {Chinni}}, \bibinfo {author} {\bibfnamefont {P.~M.}\ \bibnamefont {Poggi}},\ and\ \bibinfo {author} {\bibfnamefont {I.~H.}\ \bibnamefont {Deutsch}},\ }\bibfield  {title} {\bibinfo {title} {Effect of chaos on the simulation of quantum critical phenomena in analog quantum simulators},\ }\bibfield  {journal} {\bibinfo  {journal} {Physical Review Research}\ }\textbf {\bibinfo {volume} {3}},\ \href {https://doi.org/10.1103/physrevresearch.3.033145} {10.1103/physrevresearch.3.033145} (\bibinfo {year} {2021})\BibitemShut {NoStop}%
\bibitem [{\citenamefont {Marcantoni}\ \emph {et~al.}(2022)\citenamefont {Marcantoni}, \citenamefont {Carollo}, \citenamefont {Gambetta}, \citenamefont {Lesanovsky}, \citenamefont {Schneider},\ and\ \citenamefont {Garrahan}}]{andesson}%
  \BibitemOpen
  \bibfield  {author} {\bibinfo {author} {\bibfnamefont {S.}~\bibnamefont {Marcantoni}}, \bibinfo {author} {\bibfnamefont {F.}~\bibnamefont {Carollo}}, \bibinfo {author} {\bibfnamefont {F.~M.}\ \bibnamefont {Gambetta}}, \bibinfo {author} {\bibfnamefont {I.}~\bibnamefont {Lesanovsky}}, \bibinfo {author} {\bibfnamefont {U.}~\bibnamefont {Schneider}},\ and\ \bibinfo {author} {\bibfnamefont {J.~P.}\ \bibnamefont {Garrahan}},\ }\bibfield  {title} {\bibinfo {title} {Anderson and many-body localization in the presence of spatially correlated classical noise},\ }\bibfield  {journal} {\bibinfo  {journal} {Physical Review B}\ }\textbf {\bibinfo {volume} {106}},\ \href {https://doi.org/10.1103/physrevb.106.134211} {10.1103/physrevb.106.134211} (\bibinfo {year} {2022})\BibitemShut {NoStop}%
\bibitem [{\citenamefont {Abanin}\ \emph {et~al.}(2016)\citenamefont {Abanin}, \citenamefont {De~Roeck},\ and\ \citenamefont {Huveneers}}]{mb1}%
  \BibitemOpen
  \bibfield  {author} {\bibinfo {author} {\bibfnamefont {D.~A.}\ \bibnamefont {Abanin}}, \bibinfo {author} {\bibfnamefont {W.}~\bibnamefont {De~Roeck}},\ and\ \bibinfo {author} {\bibfnamefont {F.}~\bibnamefont {Huveneers}},\ }\bibfield  {title} {\bibinfo {title} {Theory of many-body localization in periodically driven systems},\ }\href {https://doi.org/10.1016/j.aop.2016.03.010} {\bibfield  {journal} {\bibinfo  {journal} {Annals of Physics}\ }\textbf {\bibinfo {volume} {372}},\ \bibinfo {pages} {1–11} (\bibinfo {year} {2016})}\BibitemShut {NoStop}%
\bibitem [{\citenamefont {Yunger~Halpern}\ \emph {et~al.}(2019)\citenamefont {Yunger~Halpern}, \citenamefont {White}, \citenamefont {Gopalakrishnan},\ and\ \citenamefont {Refael}}]{mb2}%
  \BibitemOpen
  \bibfield  {author} {\bibinfo {author} {\bibfnamefont {N.}~\bibnamefont {Yunger~Halpern}}, \bibinfo {author} {\bibfnamefont {C.~D.}\ \bibnamefont {White}}, \bibinfo {author} {\bibfnamefont {S.}~\bibnamefont {Gopalakrishnan}},\ and\ \bibinfo {author} {\bibfnamefont {G.}~\bibnamefont {Refael}},\ }\bibfield  {title} {\bibinfo {title} {Quantum engine based on many-body localization},\ }\bibfield  {journal} {\bibinfo  {journal} {Physical Review B}\ }\textbf {\bibinfo {volume} {99}},\ \href {https://doi.org/10.1103/physrevb.99.024203} {10.1103/physrevb.99.024203} (\bibinfo {year} {2019})\BibitemShut {NoStop}%
\bibitem [{\citenamefont {Mascarenhas}\ \emph {et~al.}(2014)\citenamefont {Mascarenhas}, \citenamefont {Bragança}, \citenamefont {Dorner}, \citenamefont {França~Santos}, \citenamefont {Vedral}, \citenamefont {Modi},\ and\ \citenamefont {Goold}}]{gold-work}%
  \BibitemOpen
  \bibfield  {author} {\bibinfo {author} {\bibfnamefont {E.}~\bibnamefont {Mascarenhas}}, \bibinfo {author} {\bibfnamefont {H.}~\bibnamefont {Bragança}}, \bibinfo {author} {\bibfnamefont {R.}~\bibnamefont {Dorner}}, \bibinfo {author} {\bibfnamefont {M.}~\bibnamefont {França~Santos}}, \bibinfo {author} {\bibfnamefont {V.}~\bibnamefont {Vedral}}, \bibinfo {author} {\bibfnamefont {K.}~\bibnamefont {Modi}},\ and\ \bibinfo {author} {\bibfnamefont {J.}~\bibnamefont {Goold}},\ }\bibfield  {title} {\bibinfo {title} {Work and quantum phase transitions: Quantum latency},\ }\bibfield  {journal} {\bibinfo  {journal} {Physical Review E}\ }\textbf {\bibinfo {volume} {89}},\ \href {https://doi.org/10.1103/physreve.89.062103} {10.1103/physreve.89.062103} (\bibinfo {year} {2014})\BibitemShut {NoStop}%
\bibitem [{\citenamefont {Campbell}(2016)}]{Campbell}%
  \BibitemOpen
  \bibfield  {author} {\bibinfo {author} {\bibfnamefont {S.}~\bibnamefont {Campbell}},\ }\bibfield  {title} {\bibinfo {title} {Criticality revealed through quench dynamics in the lipkin-meshkov-glick model},\ }\bibfield  {journal} {\bibinfo  {journal} {Physical Review B}\ }\textbf {\bibinfo {volume} {94}},\ \href {https://doi.org/10.1103/physrevb.94.184403} {10.1103/physrevb.94.184403} (\bibinfo {year} {2016})\BibitemShut {NoStop}%
\bibitem [{\citenamefont {Bento}\ \emph {et~al.}(2024)\citenamefont {Bento}, \citenamefont {del Campo},\ and\ \citenamefont {C{\'e}leri}}]{Bento2024}%
  \BibitemOpen
  \bibfield  {author} {\bibinfo {author} {\bibfnamefont {P.~H.~S.}\ \bibnamefont {Bento}}, \bibinfo {author} {\bibfnamefont {A.}~\bibnamefont {del Campo}},\ and\ \bibinfo {author} {\bibfnamefont {L.~C.}\ \bibnamefont {C{\'e}leri}},\ }\bibfield  {title} {\bibinfo {title} {Krylov complexity and dynamical phase transition in the quenched lipkin-meshkov-glick model},\ }\href@noop {} {\bibfield  {journal} {\bibinfo  {journal} {Phys. Rev. B}\ }\textbf {\bibinfo {volume} {109}},\ \bibinfo {pages} {224304} (\bibinfo {year} {2024})}\BibitemShut {NoStop}%
\bibitem [{\citenamefont {Nascimento}\ and\ \citenamefont {C{\'e}leri}(2024)}]{Nascimento2024}%
  \BibitemOpen
  \bibfield  {author} {\bibinfo {author} {\bibfnamefont {A.~B.}\ \bibnamefont {Nascimento}}\ and\ \bibinfo {author} {\bibfnamefont {L.~C.}\ \bibnamefont {C{\'e}leri}},\ }\bibfield  {title} {\bibinfo {title} {Quantum dynamical criticality speeds up thermodynamic entropy production},\ }\href@noop {} {\bibfield  {journal} {\bibinfo  {journal} {arXiv preprint arXiv:2407.03315}\ } (\bibinfo {year} {2024})}\BibitemShut {NoStop}%
\bibitem [{\citenamefont {Polkovnikov}(2008)}]{Polkovnikov2008}%
  \BibitemOpen
  \bibfield  {author} {\bibinfo {author} {\bibfnamefont {A.}~\bibnamefont {Polkovnikov}},\ }\bibfield  {title} {\bibinfo {title} {Microscopic expression for heat in the adiabatic basis},\ }\href@noop {} {\bibfield  {journal} {\bibinfo  {journal} {Physical Review Letters}\ }\textbf {\bibinfo {volume} {101}},\ \bibinfo {pages} {220402} (\bibinfo {year} {2008})}\BibitemShut {NoStop}%
\bibitem [{\citenamefont {de~Oliveira}\ and\ \citenamefont {Céleri}(2024)}]{gustavo}%
  \BibitemOpen
  \bibfield  {author} {\bibinfo {author} {\bibfnamefont {G.}~\bibnamefont {de~Oliveira}}\ and\ \bibinfo {author} {\bibfnamefont {L.~C.}\ \bibnamefont {Céleri}},\ }\bibfield  {title} {\bibinfo {title} {Thermodynamic entropy production in the dynamical casimir effect},\ }\bibfield  {journal} {\bibinfo  {journal} {Physical Review A}\ }\textbf {\bibinfo {volume} {109}},\ \href {https://doi.org/10.1103/physreva.109.012807} {10.1103/physreva.109.012807} (\bibinfo {year} {2024})\BibitemShut {NoStop}%
\bibitem [{\citenamefont {Marvian}\ and\ \citenamefont {Spekkens}(2014)}]{marvian}%
  \BibitemOpen
  \bibfield  {author} {\bibinfo {author} {\bibfnamefont {I.}~\bibnamefont {Marvian}}\ and\ \bibinfo {author} {\bibfnamefont {R.~W.}\ \bibnamefont {Spekkens}},\ }\bibfield  {title} {\bibinfo {title} {Extending noether’s theorem by quantifying the asymmetry of quantum states},\ }\bibfield  {journal} {\bibinfo  {journal} {Nature Communications}\ }\textbf {\bibinfo {volume} {5}},\ \href {https://doi.org/10.1038/ncomms4821} {10.1038/ncomms4821} (\bibinfo {year} {2014})\BibitemShut {NoStop}%
\bibitem [{\citenamefont {Sachdev}(2011)}]{sachdev2011quantum}%
  \BibitemOpen
  \bibfield  {author} {\bibinfo {author} {\bibfnamefont {S.}~\bibnamefont {Sachdev}},\ }\href@noop {} {\emph {\bibinfo {title} {Quantum phase transitions}}}\ (\bibinfo  {publisher} {Cambridge University Press},\ \bibinfo {year} {2011})\BibitemShut {NoStop}%
\bibitem [{\citenamefont {Defenu}\ \emph {et~al.}(2024)\citenamefont {Defenu}, \citenamefont {Lerose},\ and\ \citenamefont {Pappalardi}}]{pappalardi}%
  \BibitemOpen
  \bibfield  {author} {\bibinfo {author} {\bibfnamefont {N.}~\bibnamefont {Defenu}}, \bibinfo {author} {\bibfnamefont {A.}~\bibnamefont {Lerose}},\ and\ \bibinfo {author} {\bibfnamefont {S.}~\bibnamefont {Pappalardi}},\ }\bibfield  {title} {\bibinfo {title} {Out-of-equilibrium dynamics of quantum many-body systems with long-range interactions},\ }\href {https://doi.org/10.1016/j.physrep.2024.04.005} {\bibfield  {journal} {\bibinfo  {journal} {Physics Reports}\ }\textbf {\bibinfo {volume} {1074}},\ \bibinfo {pages} {1–92} (\bibinfo {year} {2024})}\BibitemShut {NoStop}%
\bibitem [{\citenamefont {Plastina}\ \emph {et~al.}(2014)\citenamefont {Plastina}, \citenamefont {Alecce}, \citenamefont {Apollaro}, \citenamefont {Falcone}, \citenamefont {Francica}, \citenamefont {Galve}, \citenamefont {Lo~Gullo},\ and\ \citenamefont {Zambrini}}]{inner}%
  \BibitemOpen
  \bibfield  {author} {\bibinfo {author} {\bibfnamefont {F.}~\bibnamefont {Plastina}}, \bibinfo {author} {\bibfnamefont {A.}~\bibnamefont {Alecce}}, \bibinfo {author} {\bibfnamefont {T.~J.~G.}\ \bibnamefont {Apollaro}}, \bibinfo {author} {\bibfnamefont {G.}~\bibnamefont {Falcone}}, \bibinfo {author} {\bibfnamefont {G.}~\bibnamefont {Francica}}, \bibinfo {author} {\bibfnamefont {F.}~\bibnamefont {Galve}}, \bibinfo {author} {\bibfnamefont {N.}~\bibnamefont {Lo~Gullo}},\ and\ \bibinfo {author} {\bibfnamefont {R.}~\bibnamefont {Zambrini}},\ }\bibfield  {title} {\bibinfo {title} {Irreversible work and inner friction in quantum thermodynamic processes},\ }\bibfield  {journal} {\bibinfo  {journal} {Physical Review Letters}\ }\textbf {\bibinfo {volume} {113}},\ \href {https://doi.org/10.1103/physrevlett.113.260601} {10.1103/physrevlett.113.260601} (\bibinfo {year} {2014})\BibitemShut {NoStop}%
\bibitem [{\citenamefont {Varizi}\ \emph {et~al.}(2022)\citenamefont {Varizi}, \citenamefont {Drumond},\ and\ \citenamefont {Landi}}]{quantum-quench}%
  \BibitemOpen
  \bibfield  {author} {\bibinfo {author} {\bibfnamefont {A.~D.}\ \bibnamefont {Varizi}}, \bibinfo {author} {\bibfnamefont {R.~C.}\ \bibnamefont {Drumond}},\ and\ \bibinfo {author} {\bibfnamefont {G.~T.}\ \bibnamefont {Landi}},\ }\bibfield  {title} {\bibinfo {title} {Quantum quench thermodynamics at high temperatures},\ }\bibfield  {journal} {\bibinfo  {journal} {Physical Review A}\ }\textbf {\bibinfo {volume} {105}},\ \href {https://doi.org/10.1103/physreva.105.062218} {10.1103/physreva.105.062218} (\bibinfo {year} {2022})\BibitemShut {NoStop}%
\bibitem [{\citenamefont {Feynman}(1939)}]{Feynman}%
  \BibitemOpen
  \bibfield  {author} {\bibinfo {author} {\bibfnamefont {R.~P.}\ \bibnamefont {Feynman}},\ }\bibfield  {title} {\bibinfo {title} {Forces in molecules},\ }\href {https://doi.org/10.1103/PhysRev.56.340} {\bibfield  {journal} {\bibinfo  {journal} {Phys. Rev.}\ }\textbf {\bibinfo {volume} {56}},\ \bibinfo {pages} {340} (\bibinfo {year} {1939})}\BibitemShut {NoStop}%
\bibitem [{\citenamefont {Lipkin}\ \emph {et~al.}(1965)\citenamefont {Lipkin}, \citenamefont {Meshkov},\ and\ \citenamefont {Glick}}]{lmg1}%
  \BibitemOpen
  \bibfield  {author} {\bibinfo {author} {\bibfnamefont {H.}~\bibnamefont {Lipkin}}, \bibinfo {author} {\bibfnamefont {N.}~\bibnamefont {Meshkov}},\ and\ \bibinfo {author} {\bibfnamefont {A.}~\bibnamefont {Glick}},\ }\bibfield  {title} {\bibinfo {title} {Validity of many-body approximation methods for a solvable model: (i). exact solutions and perturbation theory},\ }\href@noop {} {\bibfield  {journal} {\bibinfo  {journal} {Nuclear Physics}\ }\textbf {\bibinfo {volume} {62}},\ \bibinfo {pages} {188} (\bibinfo {year} {1965})}\BibitemShut {NoStop}%
\bibitem [{\citenamefont {Meshkov}\ \emph {et~al.}(1965)\citenamefont {Meshkov}, \citenamefont {Glick},\ and\ \citenamefont {Lipkin}}]{lmg2}%
  \BibitemOpen
  \bibfield  {author} {\bibinfo {author} {\bibfnamefont {N.}~\bibnamefont {Meshkov}}, \bibinfo {author} {\bibfnamefont {A.}~\bibnamefont {Glick}},\ and\ \bibinfo {author} {\bibfnamefont {H.}~\bibnamefont {Lipkin}},\ }\bibfield  {title} {\bibinfo {title} {Validity of many-body approximation methods for a solvable model: (ii). linearization procedures},\ }\href@noop {} {\bibfield  {journal} {\bibinfo  {journal} {Nuclear Physics}\ }\textbf {\bibinfo {volume} {62}},\ \bibinfo {pages} {199} (\bibinfo {year} {1965})}\BibitemShut {NoStop}%
\bibitem [{\citenamefont {Glick}\ \emph {et~al.}(1965)\citenamefont {Glick}, \citenamefont {Lipkin},\ and\ \citenamefont {Meshkov}}]{lmg3}%
  \BibitemOpen
  \bibfield  {author} {\bibinfo {author} {\bibfnamefont {A.}~\bibnamefont {Glick}}, \bibinfo {author} {\bibfnamefont {H.}~\bibnamefont {Lipkin}},\ and\ \bibinfo {author} {\bibfnamefont {N.}~\bibnamefont {Meshkov}},\ }\bibfield  {title} {\bibinfo {title} {Validity of many-body approximation methods for a solvable model: (iii). diagram summations},\ }\href@noop {} {\bibfield  {journal} {\bibinfo  {journal} {Nuclear Physics}\ }\textbf {\bibinfo {volume} {62}},\ \bibinfo {pages} {211} (\bibinfo {year} {1965})}\BibitemShut {NoStop}%
\bibitem [{\citenamefont {Romero}\ \emph {et~al.}(2022)\citenamefont {Romero}, \citenamefont {Engel}, \citenamefont {Tang},\ and\ \citenamefont {Economou}}]{lmg-nuclear}%
  \BibitemOpen
  \bibfield  {author} {\bibinfo {author} {\bibfnamefont {A.~M.}\ \bibnamefont {Romero}}, \bibinfo {author} {\bibfnamefont {J.}~\bibnamefont {Engel}}, \bibinfo {author} {\bibfnamefont {H.~L.}\ \bibnamefont {Tang}},\ and\ \bibinfo {author} {\bibfnamefont {S.~E.}\ \bibnamefont {Economou}},\ }\bibfield  {title} {\bibinfo {title} {Solving nuclear structure problems with the adaptive variational quantum algorithm},\ }\bibfield  {journal} {\bibinfo  {journal} {Physical Review C}\ }\textbf {\bibinfo {volume} {105}},\ \href {https://doi.org/10.1103/physrevc.105.064317} {10.1103/physrevc.105.064317} (\bibinfo {year} {2022})\BibitemShut {NoStop}%
\bibitem [{\citenamefont {Chen}\ \emph {et~al.}(2009)\citenamefont {Chen}, \citenamefont {Liang},\ and\ \citenamefont {Jia}}]{lmg-optica}%
  \BibitemOpen
  \bibfield  {author} {\bibinfo {author} {\bibfnamefont {G.}~\bibnamefont {Chen}}, \bibinfo {author} {\bibfnamefont {J.~Q.}\ \bibnamefont {Liang}},\ and\ \bibinfo {author} {\bibfnamefont {S.}~\bibnamefont {Jia}},\ }\bibfield  {title} {\bibinfo {title} {Interaction-induced lipkin-meshkov-glick model in a bose-einstein condensate inside an optical cavity},\ }\href {https://doi.org/10.1364/OE.17.019682} {\bibfield  {journal} {\bibinfo  {journal} {Opt. Express}\ }\textbf {\bibinfo {volume} {17}},\ \bibinfo {pages} {19682} (\bibinfo {year} {2009})}\BibitemShut {NoStop}%
\bibitem [{\citenamefont {Afrasiar}\ \emph {et~al.}(2023)\citenamefont {Afrasiar}, \citenamefont {Basak}, \citenamefont {Dey}, \citenamefont {Pal},\ and\ \citenamefont {Pal}}]{lmg-spread}%
  \BibitemOpen
  \bibfield  {author} {\bibinfo {author} {\bibfnamefont {M.}~\bibnamefont {Afrasiar}}, \bibinfo {author} {\bibfnamefont {J.~K.}\ \bibnamefont {Basak}}, \bibinfo {author} {\bibfnamefont {B.}~\bibnamefont {Dey}}, \bibinfo {author} {\bibfnamefont {K.}~\bibnamefont {Pal}},\ and\ \bibinfo {author} {\bibfnamefont {K.}~\bibnamefont {Pal}},\ }\bibfield  {title} {\bibinfo {title} {Time evolution of spread complexity in quenched lipkin–meshkov–glick model},\ }\href {https://doi.org/10.1088/1742-5468/ad0032} {\bibfield  {journal} {\bibinfo  {journal} {Journal of Statistical Mechanics: Theory and Experiment}\ }\textbf {\bibinfo {volume} {2023}},\ \bibinfo {pages} {103101} (\bibinfo {year} {2023})}\BibitemShut {NoStop}%
\bibitem [{\citenamefont {Kwok}\ \emph {et~al.}(2008)\citenamefont {Kwok}, \citenamefont {Ning}, \citenamefont {Gu},\ and\ \citenamefont {Lin}}]{lmg-condensed}%
  \BibitemOpen
  \bibfield  {author} {\bibinfo {author} {\bibfnamefont {H.-M.}\ \bibnamefont {Kwok}}, \bibinfo {author} {\bibfnamefont {W.-Q.}\ \bibnamefont {Ning}}, \bibinfo {author} {\bibfnamefont {S.-J.}\ \bibnamefont {Gu}},\ and\ \bibinfo {author} {\bibfnamefont {H.-Q.}\ \bibnamefont {Lin}},\ }\bibfield  {title} {\bibinfo {title} {Quantum criticality of the lipkin-meshkov-glick model in terms of fidelity susceptibility},\ }\bibfield  {journal} {\bibinfo  {journal} {Physical Review E}\ }\textbf {\bibinfo {volume} {78}},\ \href {https://doi.org/10.1103/physreve.78.032103} {10.1103/physreve.78.032103} (\bibinfo {year} {2008})\BibitemShut {NoStop}%
\bibitem [{\citenamefont {Heyl}(2018)}]{Heyl2018}%
  \BibitemOpen
  \bibfield  {author} {\bibinfo {author} {\bibfnamefont {M.}~\bibnamefont {Heyl}},\ }\bibfield  {title} {\bibinfo {title} {Dynamical quantum phase transitions: a review},\ }\href@noop {} {\bibfield  {journal} {\bibinfo  {journal} {Reports on Progress in Physics}\ }\textbf {\bibinfo {volume} {81}},\ \bibinfo {pages} {054001} (\bibinfo {year} {2018})}\BibitemShut {NoStop}%
\bibitem [{\citenamefont {Nachbin}\ and\ \citenamefont {Bechtolsheim}(1965)}]{nachbin}%
  \BibitemOpen
  \bibfield  {author} {\bibinfo {author} {\bibfnamefont {L.}~\bibnamefont {Nachbin}}\ and\ \bibinfo {author} {\bibfnamefont {L.}~\bibnamefont {Bechtolsheim}},\ }\href {https://books.google.com.br/books?id=KMA-AAAAIAAJ} {\emph {\bibinfo {title} {The Haar Integral}}},\ University Series in Higher Mathematics : a series of advanced text and reference books in pure and applied mathematics\ (\bibinfo  {publisher} {Van Nostrand},\ \bibinfo {year} {1965})\BibitemShut {NoStop}%
\bibitem [{\citenamefont {Martin}(2021)}]{sanmartin-grupodelie}%
  \BibitemOpen
  \bibfield  {author} {\bibinfo {author} {\bibfnamefont {L.}~\bibnamefont {Martin}},\ }\href {https://books.google.com.br/books?id=llKvswEACAAJ} {\emph {\bibinfo {title} {Grupos De Lie}}}\ (\bibinfo  {publisher} {UNICAMP},\ \bibinfo {year} {2021})\BibitemShut {NoStop}%
\bibitem [{\citenamefont {Halmos}(1974)}]{Fubbini-Halmos}%
  \BibitemOpen
  \bibfield  {author} {\bibinfo {author} {\bibfnamefont {P.~R.}\ \bibnamefont {Halmos}},\ }\href@noop {} {\emph {\bibinfo {title} {Measure Theory}}}\ (\bibinfo  {publisher} {Springer Verlag},\ \bibinfo {year} {1974})\BibitemShut {NoStop}%
\bibitem [{\citenamefont {Barnum}\ \emph {et~al.}(1998)\citenamefont {Barnum}, \citenamefont {Nielsen},\ and\ \citenamefont {Schumacher}}]{nielsen-channel}%
  \BibitemOpen
  \bibfield  {author} {\bibinfo {author} {\bibfnamefont {H.}~\bibnamefont {Barnum}}, \bibinfo {author} {\bibfnamefont {M.~A.}\ \bibnamefont {Nielsen}},\ and\ \bibinfo {author} {\bibfnamefont {B.}~\bibnamefont {Schumacher}},\ }\bibfield  {title} {\bibinfo {title} {Information transmission through a noisy quantum channel},\ }\href {https://doi.org/10.1103/physreva.57.4153} {\bibfield  {journal} {\bibinfo  {journal} {Physical Review A}\ }\textbf {\bibinfo {volume} {57}},\ \bibinfo {pages} {4153–4175} (\bibinfo {year} {1998})}\BibitemShut {NoStop}%
\bibitem [{\citenamefont {Cover}\ and\ \citenamefont {Thomas}(2012)}]{cover}%
  \BibitemOpen
  \bibfield  {author} {\bibinfo {author} {\bibfnamefont {T.}~\bibnamefont {Cover}}\ and\ \bibinfo {author} {\bibfnamefont {J.}~\bibnamefont {Thomas}},\ }\href {https://books.google.com.br/books?id=VWq5GG6ycxMC} {\emph {\bibinfo {title} {Elements of Information Theory}}}\ (\bibinfo  {publisher} {Wiley},\ \bibinfo {year} {2012})\BibitemShut {NoStop}%
\bibitem [{\citenamefont {Wilde}(2016)}]{wilde}%
  \BibitemOpen
  \bibfield  {author} {\bibinfo {author} {\bibfnamefont {M.~M.}\ \bibnamefont {Wilde}},\ }\bibinfo {title} {Quantum information theory}\ (\bibinfo  {publisher} {Cambridge University Press},\ \bibinfo {year} {2016})\ p.\ \bibinfo {pages} {xi–xii}\BibitemShut {NoStop}%
\end{thebibliography}%

%

\end{document}